\PassOptionsToPackage{unicode}{hyperref}
\PassOptionsToPackage{hyphens}{url}
\documentclass[11pt]{article}
\usepackage{xcolor}
\usepackage{amsmath,amssymb}
\usepackage{graphicx}
\usepackage{subcaption}
\usepackage{float}
\usepackage{geometry}
\usepackage{cite}
\usepackage{tikz}
\usepackage{siunitx}
\usetikzlibrary{arrows.meta,positioning,calc}
\usepackage{iftex}
\ifPDFTeX
  \usepackage[T1]{fontenc}
  \usepackage[utf8]{inputenc}
  \usepackage{textcomp} 
\else 
  \usepackage{unicode-math} 
  \defaultfontfeatures{Scale=MatchLowercase}
  \defaultfontfeatures[\rmfamily]{Ligatures=TeX,Scale=1}
\fi
\usepackage{lmodern}
\ifPDFTeX\else
\fi
\IfFileExists{upquote.sty}{\usepackage{upquote}}{}
\IfFileExists{microtype.sty}{
  \usepackage[]{microtype}
  \UseMicrotypeSet[protrusion]{basicmath} 
}{}
\makeatletter
\@ifundefined{KOMAClassName}{
  \IfFileExists{parskip.sty}{%
    \usepackage{parskip}
  }{
    \setlength{\parindent}{0pt}
    \setlength{\parskip}{6pt plus 2pt minus 1pt}}
}{
  \KOMAoptions{parskip=half}}
\makeatother
\usepackage{longtable,booktabs,array}
\usepackage{calc} 
\usepackage{etoolbox}
\makeatletter
\patchcmd\longtable{\par}{\if@noskipsec\mbox{}\fi\par}{}{}
\makeatother
\IfFileExists{footnotehyper.sty}{\usepackage{footnotehyper}}{\usepackage{footnote}}
\makesavenoteenv{longtable}
\usepackage{bookmark}
\IfFileExists{xurl.sty}{\usepackage{xurl}}{} 
\graphicspath{{figures/images/}}
\hypersetup{
  hidelinks,
  pdfcreator={LaTeX via pandoc}}

\title{Probabilistic Frequency Hazard Analysis: Adapting the Seismic Hazard Framework to Power System Frequency Exceedance Risk}
\author{Sewedo Todowede \\ Frequency Risk and Modelling Team \\ National Energy System Operator (NESO), Great Britain \\ \texttt{sewedo.todowede@neso.energy}}
\date{}

\begin{document}

\maketitle

\noindent\footnotesize\emph{Disclaimer: This paper is an individual technical research contribution. The views and conclusions expressed here are those of the author alone and do not necessarily represent the official position of the National Energy System Operator (NESO). The PFHA framework described here is not an operational NESO tool.}\normalsize

\vspace{0.5em}

\begin{abstract}

The declining synchronous inertia in power systems undergoing the energy
transition increases the sensitivity of system frequency to generation
and interconnector disturbances, making accurate frequency risk
quantification increasingly important. Existing methods for frequency
risk assessment, while valuable, lack formal uncertainty quantification,
continuous hazard curves, and source-level disaggregation. This paper
introduces Probabilistic Frequency Hazard Analysis (PFHA), a framework
that adapts the mathematical architecture of Probabilistic Seismic
Hazard Analysis (PSHA), the standard methodology in earthquake
engineering, to power system frequency
exceedance risk. The PFHA hazard integral computes annual exceedance
rates by integrating over all combinations of loss sources, disturbance
sizes, and system operating states through a frequency response
prediction equation with calibrated aleatory variability. The framework
is implemented with a 51-source catalogue constructed from operational
data, empirical loss distributions from settlement-period generation
records, Bayesian occurrence rate estimation, a dual analytical and
physics-based frequency response prediction architecture, and a 324-path
logic tree for epistemic uncertainty quantification. Application to the
Great Britain power system using four years of operational data
demonstrates agreement with the independently developed Frequency Risk
and Control Report to within a factor of 1.5 at 49.2 Hz, while also
quantifying the risk reduction from Dynamic Containment and
Low-Frequency Demand Disconnection controls. To the author's knowledge,
this is the first published explicit PSHA-style hazard-integral
formulation for bulk power-system frequency exceedance risk.
\end{abstract}


\subsection*{Nomenclature}
\begin{tabular}{@{}ll@{}}
$\lambda$ & Annual frequency exceedance rate (events/yr) \\
$\delta$ & Frequency deviation threshold (Hz below $f_0$) \\
$\Delta f$ & Frequency nadir deviation (Hz) \\
$\Delta P$ & Power imbalance / loss magnitude (MW) \\
$\nu_i$ & Annual trip rate of source $i$ (events/yr) \\
$f_i(\Delta P)$ & Loss-size probability mass function of source $i$ \\
$g(\mathbf{s})$ & Joint probability density of system state \\
$\mathbf{s}$ & System state vector $(H, D, R)$ \\
$H$ & System inertia (GVA$\cdot$s) \\
$D$ & System demand (GW) \\
$R$ & Frequency response holdings (MW) \\
$\mu$ & Median frequency nadir predicted by FRPE (Hz) \\
$\sigma$ & Aleatory variability (log-space standard deviation) \\
$b$ & Bias correction factor (SFR paths only) \\
$\Phi$ & Standard normal cumulative distribution function \\
$\varepsilon$ & Epsilon: number of $\sigma$ above median in log-space \\
$f_0$ & Nominal system frequency (50 Hz) \\
$M$ & Effective rotational inertia ($= 2H \times 1000/f_0$, MW$\cdot$s/Hz, with $H$ in GVA$\cdot$s) \\
$D_{\mathrm{eff}}$ & Total effective damping (MW/Hz) \\
$w_j, w_k$ & PMF bin weight, state bin weight \\
$N_{\mathrm{src}}$ & Number of sources \\
$N_{\mathrm{bin}}$ & Number of empirical state bins \\
\end{tabular}

\section{I. Introduction}\label{i.-introduction}

\subsection{A. The Frequency Risk
Challenge}\label{a.-the-frequency-risk-challenge}

The progressive displacement of synchronous generation by
converter-interfaced renewable sources is altering the
frequency dynamics of power systems worldwide. In the Great Britain (GB)
system, aggregate system inertia has declined from typical levels above 300
GVA\(\cdot\)s in 2010 \cite{ngeso_inertia_naspi_2021} to an approved minimum operating level of 120 GVA\(\cdot\)s \cite{neso_frcr_2024}. Lower inertia increases the rate
of change of frequency (RoCoF) for a given power imbalance and deepens
the frequency nadir, amplifying the consequences of generation and
interconnector loss events that would previously have been absorbed with
minimal frequency deviation.

The significance of this shift was demonstrated by the event of 9 August
2019, when the near-simultaneous loss of the Hornsea 1 offshore wind
farm and Little Barford gas turbine, compounded by distributed energy
resource disconnection triggered by high RoCoF, produced a total power
deficit of approximately 1700 MW. System frequency fell to 48.8 Hz,
the threshold at which the first stage of Low-Frequency Demand
Disconnection (LFDD) activates, and approximately one million
consumers were disconnected \cite{ngeso_2019_aug9}. This event occurred at a system
inertia of approximately 210 GVA\(\cdot\)s. The system now routinely operates below 140 GVA\(\cdot\)s during high-wind periods, where the same compound loss would produce a substantially deeper frequency excursion.

The GB system manages frequency risk through the annual Frequency Risk
and Control Report (FRCR), published by the National Energy System
Operator (NESO). The FRCR uses settlement-period-resolution simulation
sampling system conditions from operational distributions to estimate frequency deviation probabilities and to
determine the cost-optimal balance of response holdings and minimum
inertia constraints \cite{neso_frcr_2024}. It represents a significant and pioneering
effort in operational frequency risk management.

\subsection{B. Motivation for a Complementary Framework}\label{b.-the-auditability-gap}

An independent consultant review commissioned by the Office of Gas and
Electricity Markets (Ofgem) of the FRCR 2025 submission identified
opportunities for enhancing the transparency of probability estimation
methodology, the formal validation of simulations against empirical
observations, and the auditability of reported risk figures
\cite{ofgem_frcr_2025_review}. These observations relate to the
presentation of evidence and assumptions rather than to the underlying
principles or overall results of the FRCR methodology.

More broadly, the deterministic settlement-period enumeration approach, while
computationally established, does not naturally produce several outputs
that probabilistic hazard analysis in other domains considers standard:
continuous hazard curves expressing exceedance rate as a function of
threshold, formal separation and quantification of epistemic and
aleatory uncertainty, source-level disaggregation identifying which
generation or interconnector sources dominate risk at each severity
level, and systematic quantification of risk reduction attributable to
individual control measures. These considerations motivate the
development of a complementary analytical framework grounded in
established probabilistic hazard methods.

\subsection{C. Related Work}\label{c.-related-work}

Existing approaches to frequency risk quantification fall into three categories: operational risk frameworks maintained by system operators, probabilistic frequency stability assessment methods in the academic literature, and probabilistic hazard frameworks from other engineering domains.

Among operational frameworks, the GB FRCR \cite{neso_frcr_2024} uses deterministic settlement-period enumeration to estimate frequency deviation probabilities at three regulatory thresholds, and is the primary benchmark against which PFHA is validated. In Australia, AEMO published a dedicated Power System Frequency Risk Review \cite{aemo_psfrr_2022} assessing frequency risks from non-credible contingency events in the National Electricity Market; this has since been incorporated into the broader General Power System Risk Review \cite{aemo_gpsrr_2025}. In Continental Europe, ENTSO-E's Project Inertia studies \cite{entsoe_inertia_2021,entsoe_inertia_2023} assess frequency stability under future scenarios, focusing on system-split events and RoCoF thresholds.

In the academic literature, probabilistic approaches to frequency nadir assessment have emerged. Wen \emph{et al.} \cite{wen_bu_xin_2021} propose a probabilistic assessment of area-level frequency nadir for operational planning, using a multi-interval sensitivity method to compute the probabilistic distribution of frequency nadir under renewable energy uncertainty. Milanovi\'{c} \cite{milanovic_2017} presents a probabilistic stability analysis framework arguing that probabilistic methods are necessary for sustainable power systems with high renewable penetration. Shahzad \cite{shahzad_2022} develops a probabilistic framework for large-disturbance global instability risk assessment incorporating frequency stability alongside angle and voltage stability.


From adjacent domains, Vlachopoulou \emph{et al.} \cite{vlachopoulou_2016} applied PSHA to the physical seismic risk assessment of power grid infrastructure, demonstrating the feasibility of transferring probabilistic hazard methods to power systems, although their application addressed structural fragility rather than frequency exceedance. The probabilistic reliability evaluation frameworks of Billinton and Allan \cite{billinton_allan_1996} and Li \cite{li_2005} provide the foundational probabilistic methods upon which the PFHA builds, while Kundur \cite{kundur_1994} and Milano \cite{milano_2010} provide the frequency dynamics theory underlying the FRPE models.

\subsection{D. The Seismic Hazard
Analogy}\label{d.-the-seismic-hazard-analogy}

Probabilistic Seismic Hazard Analysis (PSHA), formalised by Cornell
\cite{cornell_1968} and refined over several decades of practice \cite{baker_bradley_stafford_2021,baker_2013_psha,sshac_1997},
provides a rigorous mathematical framework for computing the rate of
exceeding ground-motion thresholds at a site given all possible
earthquake scenarios. The PSHA hazard integral sums over seismic
sources, integrating over magnitude distributions and source-to-site
distances through a ground-motion prediction equation (GMPE) that
predicts intensity with calibrated aleatory variability. Epistemic
uncertainty in model choices is captured via logic trees with weighted
branches.

The PSHA framework has been successfully transferred to other natural
hazard domains, including probabilistic tsunami hazard analysis (PTHA)
\cite{grezio_2017_ptha} and volcanic hazard analysis (PVHA) \cite{marzocchi_bebbington_2012}. Each transfer
preserves the mathematical architecture (the total probability
theorem, the separation of aleatory and epistemic uncertainty, the logic
tree, the disaggregation) while adapting the physical models to the
new domain.

The central observation motivating this paper is that power system
frequency exceedance shares the same mathematical structure as seismic
ground-motion exceedance. A generation or interconnector source produces
a disturbance (analogous to an earthquake), whose size follows a
distribution (analogous to a magnitude distribution). The frequency
response of the system is predicted by a Frequency Response Prediction
Equation (FRPE, analogous to a GMPE) that depends on system operating
conditions (analogous to site conditions) and carries aleatory
variability. The rate of exceeding a frequency threshold is the sum over
all sources, integrated over all disturbance sizes and system states,
structurally identical to the PSHA integral. Previous work has applied
elements of PSHA to physical seismic risk assessment of power grid
infrastructure \cite{vlachopoulou_2016}, but to the author's knowledge,
no previous work has formulated an explicit PSHA-style hazard integral
for power system frequency exceedance risk.

\subsection{E. Contributions and Paper
Organisation}\label{e.-contributions-and-paper-organisation}

This paper makes the following contributions:

\begin{enumerate}
\def\labelenumi{\arabic{enumi})}
\item
  Formulation of the PFHA hazard integral adapting the PSHA mathematical
  framework to frequency exceedance risk, with identification of the
  three key extensions required: non-stationarity of system state,
  correlated multi-source failures, and operator-controlled frequency
  response.
\item
  A 51-source catalogue for the GB system with empirical loss
  distributions from settlement-period operational data and Bayesian
  occurrence rate estimation that eliminates the zero-rate problem
  inherent in maximum likelihood estimation.
\item
  A dual-FRPE architecture combining an analytical system frequency
  response (SFR) model with a physics-based frequency simulation
  engine, embedded within a 324-path logic tree for epistemic
  uncertainty quantification.
\item
  Application to the GB power system demonstrating cross-validation with
  the independently-developed FRCR within a factor of 1.5 at the 49.2 Hz
  threshold.
\item
  Quantification of the risk reduction from Dynamic Containment (DC) and LFDD
  controls, a capability not available from existing methods.
\end{enumerate}

The remainder of this paper is organised as follows. Section II
formulates the PFHA hazard integral. Sections III--VI describe the four
input components: source characterisation, frequency response
prediction, system state integration, and control modelling. Section VII
presents the logic tree and uncertainty quantification framework.
Sections VIII and IX present the GB application results and validation,
and Section X discusses limitations and future work.

\section{II. The PFHA Hazard
Integral}\label{ii.-the-pfha-hazard-integral}

\subsection{A. Mathematical
Formulation}\label{a.-mathematical-formulation}

The PFHA hazard integral is derived from the total probability theorem.
The annual rate \(\lambda\) of frequency deviations exceeding a
threshold \(\delta\) is obtained by summing over all
\(N_{\mathrm{src}}\) sources,
integrating over the distribution of disturbance sizes \(\Delta P\) and
system operating states \(\mathbf{s}\):

\[\lambda(\Delta f > \delta) = \sum_{i=1}^{N_{\mathrm{src}}} \nu_i \int_{\Delta P} \int_{\mathbf{s}} P(\Delta f > \delta \mid \Delta P, \mathbf{s}) \cdot f_i(\Delta P) \cdot g(\mathbf{s}) \, d\Delta P \, d\mathbf{s} \tag{1}\]

where \(\nu_i\) is the annual trip rate of source \(i\) estimated via
Bayesian updating of technology-class priors with observed event counts
(Section III-C); \(f_i(\Delta P)\) is the empirical probability mass
function (PMF) of loss size for source \(i\), constructed from
settlement-period generation data (Section III-B); \(g(\mathbf{s})\) is
the joint probability density of the system state vector
\(\mathbf{s} = (H, D, R)\) representing inertia, demand, and frequency
response holdings, estimated from 71,476 empirical half-hourly
settlement periods (Section V); and
\(P(\Delta f > \delta \mid \Delta P, \mathbf{s})\) is the conditional
exceedance probability provided by the Frequency Response Prediction
Equation (Section IV).

Following the PSHA convention for ground-motion prediction \cite{baker_bradley_stafford_2021}, the
conditional exceedance probability is computed under a log-normal model:

\[P(\Delta f > \delta \mid \Delta P, \mathbf{s}) = \Phi\left(\frac{\ln\left(\mu_i(\Delta P, \mathbf{s})\right) - \ln(\delta)}{\sigma_i(\Delta P, \mathbf{s})}\right) \tag{2}\]

where \(\mu_i(\Delta P, \mathbf{s})\) is the median frequency nadir
predicted by the FRPE, \(\sigma_i(\Delta P, \mathbf{s})\) is the
aleatory variability (calibrated from event replay residuals), and
\(\Phi\) is the standard normal cumulative distribution function. The
argument represents the number of standard deviations by which the
median prediction exceeds the threshold in log-space.

\subsection{B. The PSHA--PFHA Component
Mapping}\label{b.-the-pshapfha-component-mapping}

Table~\ref{tab:mapping} summarises the structural correspondence between the PSHA and
PFHA frameworks. Each PSHA component maps to a PFHA counterpart with an
identifiable data source. The mapping preserves the mathematical
architecture while substituting power-system-specific physical models
and data.

\begin{table}[t]
\centering
\caption{PSHA--PFHA Component Mapping}
\label{tab:mapping}
\begin{tabular}{>{\raggedright\arraybackslash}p{3.7cm}>{\raggedright\arraybackslash}p{4.1cm}>{\raggedright\arraybackslash}p{3.7cm}}
\toprule
PSHA Component & PFHA Analogue & Data Source \\
\midrule
Seismic source (fault) & Loss source (generator/IC) & 51-source GB catalogue \\
Magnitude distribution $f(M)$ & Loss-size distribution $f_i(\Delta P)$ & B1610 empirical PMFs \\
GMPE & FRPE & Analytical SFR / Physics-based lookup \\
Occurrence rate $\nu$ (Poisson) & Trip rate $\nu_i$ (Bayesian Gamma-Poisson) & GC0105 incident reports \\
Site conditions ($V_{s30}$) & System state $\mathbf{s} = (H, D, R)$ & Half-hourly operational data \\
Logic tree (epistemic) & Logic tree (6 branches, 324 paths) & Calibration + engineering judgement \\
\bottomrule
\end{tabular}
\end{table}

\begin{figure}[t]
\centering
\begin{tikzpicture}[
  >=Latex,
  every node/.style={font=\small},
  box/.style={draw, rounded corners, align=center, minimum width=3.1cm, minimum height=0.85cm},
  maparrow/.style={-{Latex[length=2mm]}, thick, gray!70}
]
\node[box, fill=blue!6] (psha1) at (0,0) {Seismic\\source};
\node[box, fill=blue!6] (psha2) at (0,-1.4) {Magnitude\\distribution};
\node[box, fill=blue!6] (psha3) at (0,-2.8) {GMPE};
\node[box, fill=blue!6] (psha4) at (0,-4.2) {Site\\conditions};
\node[box, fill=blue!6] (psha5) at (0,-5.6) {Hazard\\curve};

\node[box, fill=green!8] (pfha1) at (6.2,0) {Loss\\source};
\node[box, fill=green!8] (pfha2) at (6.2,-1.4) {Loss PMF};
\node[box, fill=green!8] (pfha3) at (6.2,-2.8) {FRPE};
\node[box, fill=green!8] (pfha4) at (6.2,-4.2) {System\\state};
\node[box, fill=green!8] (pfha5) at (6.2,-5.6) {Hazard\\curve};

\draw[maparrow] (psha1) -- (psha2);
\draw[maparrow] (psha2) -- (psha3);
\draw[maparrow] (psha3) -- (psha4);
\draw[maparrow] (psha4) -- (psha5);

\draw[maparrow] (pfha1) -- (pfha2);
\draw[maparrow] (pfha2) -- (pfha3);
\draw[maparrow] (pfha3) -- (pfha4);
\draw[maparrow] (pfha4) -- (pfha5);

\draw[maparrow] (psha1.east) -- (pfha1.west);
\draw[maparrow] (psha2.east) -- (pfha2.west);
\draw[maparrow] (psha3.east) -- (pfha3.west);
\draw[maparrow] (psha4.east) -- (pfha4.west);
\draw[maparrow] (psha5.east) -- (pfha5.west);
\end{tikzpicture}
\caption{Structural correspondence between PSHA and PFHA. Each component in the seismic hazard framework maps to a power-system counterpart, preserving the hazard-integral architecture while substituting domain-specific physics and data.}
\label{fig:psha-pfha-mapping}
\end{figure}
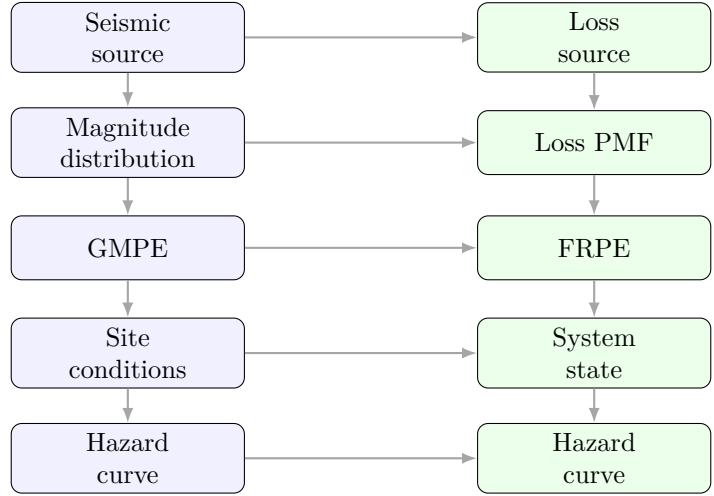

A foundational principle inherited from PSHA practice is the separation
of loss-size characterisation from occurrence rate estimation. Loss
sizes are characterised from abundant operating data (approximately
70,000 settlement periods per source), while occurrence rates are
estimated from sparse incident data via Bayesian methods. These two
concerns are mathematically independent in (1): the PMF weights
\(f_i(\Delta P)\) and the rate \(\nu_i\) enter as separate
multiplicative terms, enabling each to be estimated with the most
appropriate data and methodology.

\subsection{C. Numerical
Discretisation}\label{c.-numerical-discretisation}

The continuous integrals in (1) are approximated by discrete sums over
PMF bins and state bins:

\[\lambda(\Delta f > \delta) \approx \sum_{i=1}^{N_{\mathrm{src}}} \nu_i \sum_{j=1}^{N_{\Delta P}^{(i)}} \sum_{k=1}^{N_{\mathrm{bin}}} w_j^{(i)} \cdot w_k \cdot \Phi\left(\frac{\ln(\mu_{ijk}) - \ln(\delta)}{\sigma_{ijk}}\right) \tag{3}\]

where \(w_j^{(i)}\) is the PMF weight for the \(j\)-th loss bin of
source \(i\), \(w_k\) is the empirical weight for the \(k\)-th state
bin, and \(\mu_{ijk}\) and \(\sigma_{ijk}\) are evaluated at loss bin
\(j\) and state bin \(k\). The implementation uses
\(N_{\mathrm{bin}} = 50\)
quantile-based state bins and source-specific PMF resolutions of 25 MW
(typically 10--40 bins per source), yielding approximately 51,000
integration cells per logic tree path. Vectorised array operations
 eliminate explicit iteration over the triple sum, achieving
 approximately 1–2 seconds computation time per path.

\subsection{D. Worked Example}\label{d.-worked-example}

To illustrate the hazard integral concretely, consider a single source (Sizewell~B nuclear station, $\nu = 0.15$/yr, capacity 1198~MW) at a single system state ($H = 180$~GVA$\cdot$s, $D = 28$~GW, $R = 1500$~MW) evaluated at the 49.2~Hz threshold ($\delta = 0.8$~Hz).

The SFR model (with bias $b = 0.370$) predicts a median nadir of $\mu = 0.164$~Hz. The aleatory sigma at this point is $\sigma = 0.296 \times 1.0 \times 1.07 = 0.317$ (inertia factor = 1.0, size factor = 1.07 for 1198~MW). The z-score is:
\[z = \frac{\ln(0.164) - \ln(0.8)}{0.317} = \frac{-1.807 - (-0.223)}{0.317} = -4.99\]
The exceedance probability is $P = \Phi(-4.99) \approx 3.0 \times 10^{-7}$; the median prediction is far below the threshold, so only extreme tail events would breach it. The contribution to the 49.2~Hz rate from this single (source, state, loss-bin) cell is $0.15 \times 3.0 \times 10^{-7} \times w_j \times w_k$, which is negligible.

For the same source at the 49.5~Hz threshold ($\delta = 0.5$~Hz), $z = -3.52$, giving $P = \Phi(-3.52) \approx 2.2 \times 10^{-4}$. The contribution is larger but still modest: Sizewell~B's low trip rate (0.15/yr) limits its hazard contribution even when the exceedance probability is non-trivial.

The full hazard at each threshold is the sum of such contributions over all 51 sources, all PMF bins, and all 50 state bins, approximately 51,000 integration cells.

\subsection{E. Key Extensions Beyond Classical
PSHA}\label{e.-key-extensions-beyond-classical-psha}

Three extensions to the classical PSHA formulation are required for the
power system domain:

\begin{enumerate}
\def\labelenumi{\arabic{enumi})}
\item
  \emph{Non-stationarity:} Power system state changes on hourly
  timescales (inertia varies between 100 and 300 GVA\(\cdot\)s within a single
  day), unlike the quasi-static site conditions in seismology. This is
  handled by the empirical state integration in (1), which weights each
  operating regime by its observed frequency of occurrence.
\item
  \emph{Correlated failures:} The PSHA assumption of independent source
  events does not hold for all power system scenarios. Simultaneous pair
  events (Layer B) and compound cascade events (Layer C) are modelled
  through an explicit multi-layer source architecture (Section III-D).
\item
  \emph{Operator intervention:} Dynamic Containment and LFDD modify the
  frequency trajectory after a disturbance, a control mechanism with
  no direct analogue in PSHA. These are incorporated via a control model
  that modifies the FRPE output within the integrand (Section VI).
\end{enumerate}

\section{III. Source
Characterisation}\label{iii.-source-characterisation}

\subsection{A. Source Catalogue}\label{a.-source-catalogue}

The PFHA source catalogue comprises 51 individual sources (Layer A)
representing generation units and interconnectors on the GB system whose
trip would produce a frequency disturbance. These include 19
individually disaggregated combined-cycle gas turbines (CCGTs), 10
interconnectors (with IFA split into its two bipoles), 2 nuclear
stations, 1 biomass station, 3 pumped storage stations, 12 large wind
farms, and 2 fleet-level catch-all sources for unmatched CCGT and wind
events. Source identification is achieved by mapping Balancing Mechanism
Unit (BMU) identifiers from the Elexon BMRS reporting system to PFHA
source identifiers via a manually-curated registry.

Table~\ref{tab:catalogue} summarises the catalogue by source type.

\begin{table}[t]
\centering
\caption{Source Catalogue Summary}
\label{tab:catalogue}
\begin{tabular}{llll}
\toprule
Source Type & Count & Capacity Range (MW) & Rate Range (/yr) \\
\midrule
CCGT (individual) & 19 & 400--880 & 0.16--4.24 \\
Interconnector & 10 & 500--1400 & 0.39--13.4 \\
Nuclear & 2 & 580--1200 & 0.51--1.26 \\
Biomass & 1 & 660 & 0.76 \\
Pumped storage & 3 & 300--1800 & 0.34--0.51 \\
Wind & 12 & 400--1200 & 0.34--0.51 \\
Fleet catch-all & 2 & varies & 0.51--47.7 \\
\bottomrule
\end{tabular}
\end{table}

\subsection{B. Empirical Loss
Distributions}\label{b.-empirical-loss-distributions}

The conditional loss distribution \(f_i(\Delta P)\) for each source is
constructed as an empirical PMF from half-hourly actual generation data
(Elexon B1610 dataset). For each source,
the pipeline loads B1610 records for the source's constituent BMUs, sums
output across BMUs at each settlement period, filters to periods of
positive generation (approximately 70,000 per source for baseload
units), histograms the output into 25 MW bins, truncates at the source's
maximum credible loss, and normalises to unit probability.

This approach captures the true operational output profile, including
bimodal patterns in interconnector flows (clustering at full import and
zero) and seasonal variation in wind farm output, without parametric
assumptions.

\begin{figure}[t]
\centering
\begin{subfigure}[t]{0.48\textwidth}
  \centering
  \includegraphics[width=\linewidth]{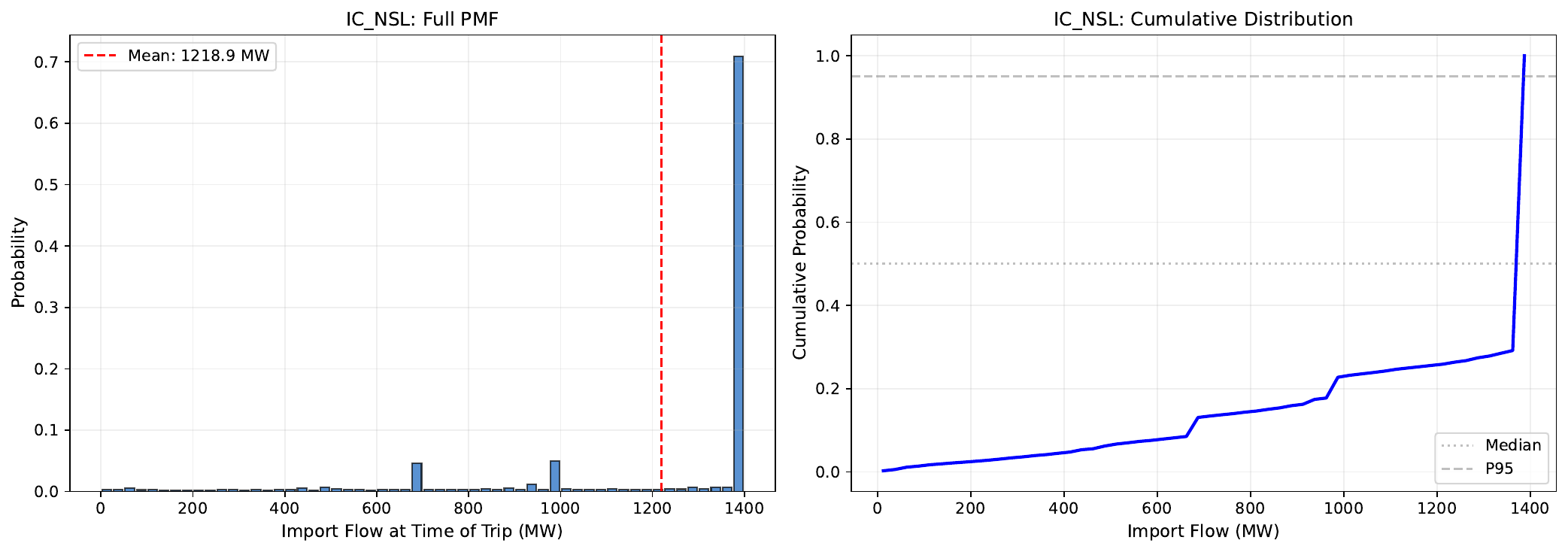}
  \caption{IC\_NSL empirical PMF.}
\end{subfigure}
\hfill
\begin{subfigure}[t]{0.48\textwidth}
  \centering
  \includegraphics[width=\linewidth]{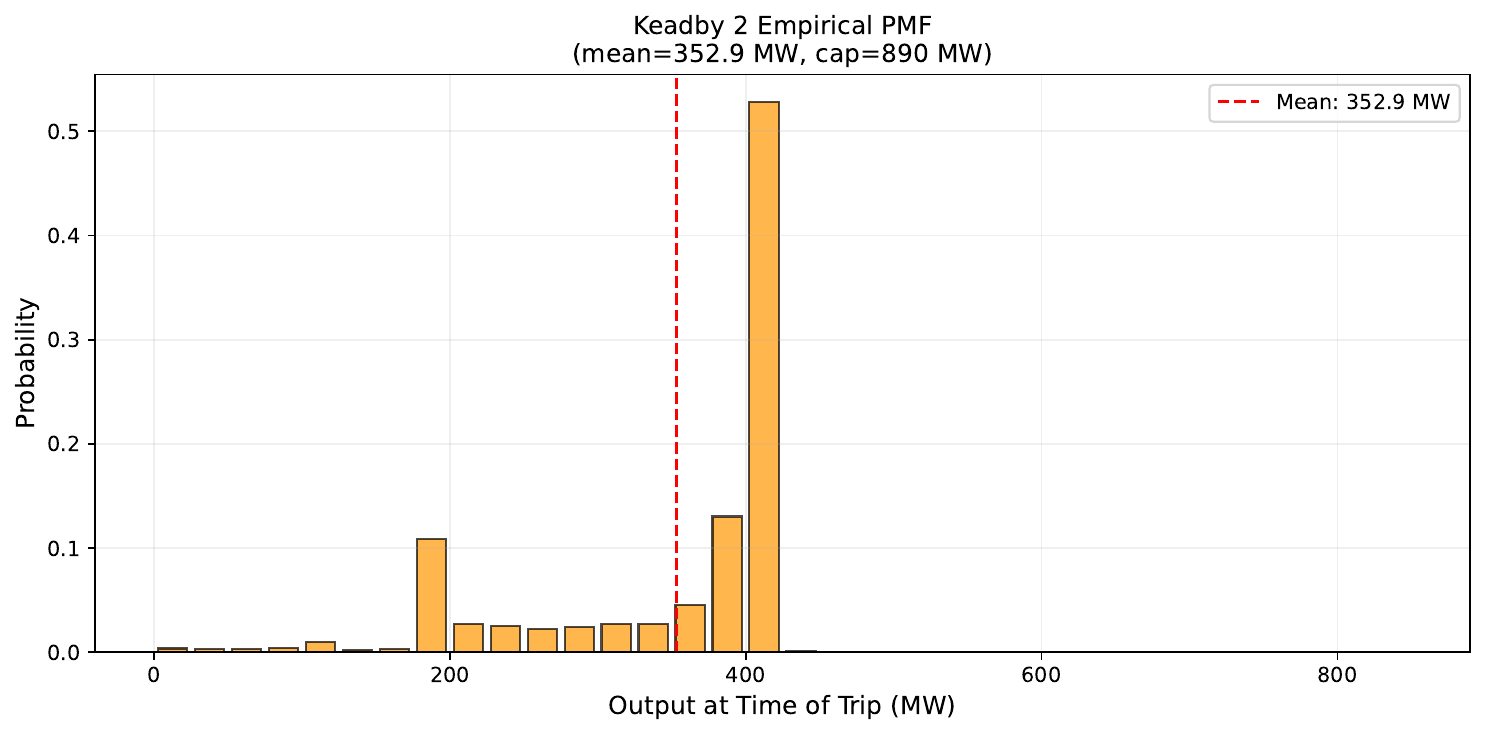}
  \caption{GEN\_CCGT\_KEAD2 empirical PMF.}
\end{subfigure}
\caption{Empirical loss-size probability mass functions constructed from 24 months of B1610 settlement-period generation data. The North Sea Link interconnector exhibits bimodal flow patterns, while Keadby~2 concentrates near rated output.}
\label{fig:example-pmfs}
\end{figure}

\subsection{C. Bayesian Rate
Estimation}\label{c.-bayesian-rate-estimation}

Maximum likelihood estimation (MLE) of trip rates,
\(\hat{\nu} = n / T\), gives zero for sources with no observed trips in
the four-year observation window. A zero rate renders the source
invisible in the hazard integral regardless of its capacity, a
physically unreasonable outcome for operational sources. The Bayesian
Gamma-Poisson conjugate framework resolves this:

\[\lambda_i \mid n_i \sim \mathrm{Gamma}(\alpha_c + n_i, \;\beta_c + T_i) \tag{4}\]

where \(\alpha_c\) and \(\beta_c\) are technology-class prior parameters
(six classes: CCGT, nuclear, biomass, interconnector, wind, pumped
storage), \(n_i\) is the count of observed trips from GC0105 incident
reports, and \(T_i\) is the source-specific observation period.
The posterior mean \((\alpha_c + n_i) / (\beta_c + T_i)\) is a weighted
average of the prior mean and the MLE: when \(n_i\) is large (e.g., IFA
with 53 observed trips), the posterior is data-dominated; when
\(n_i = 0\) (e.g., wind farms), the posterior is prior-dominated but
remains non-zero.

GC0105 events that could not be matched to
individually-catalogued sources are assigned to the CCGT fleet catch-all
at an aggregate rate of 47.7/yr. Sensitivity analysis (Section VII-D)
confirms this assignment is non-material to the results at
decision-relevant thresholds.

\begin{figure}[t]
\centering
\includegraphics[width=0.8\textwidth]{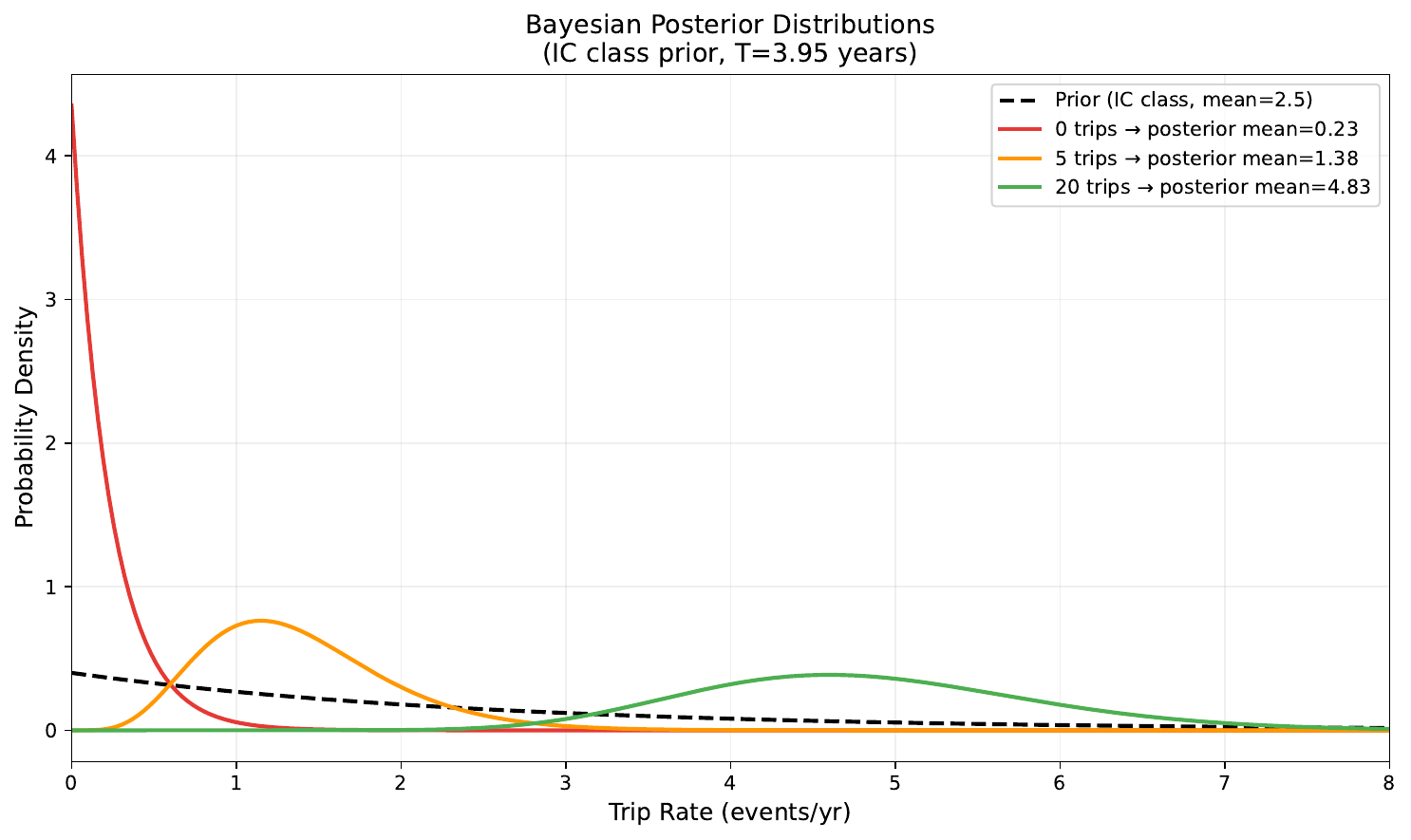}
\caption{Bayesian updating of source trip rates under the Gamma--Poisson conjugate model. With zero observed events the posterior remains prior-dominated, while abundant data drive the posterior toward the maximum-likelihood estimate.}
\label{fig:bayesian-updating}
\end{figure}

\subsection{D. Multi-Layer Source
Model}\label{d.-multi-layer-source-model}

Beyond individual source trips (Layer A), the PFHA models two additional
event classes that dominate the deep-threshold hazard.

\emph{1) Layer B --- Simultaneous Pairs:} The 30 Layer B pairs represent
credible combinations of two large sources tripping within the same
settlement period, producing combined losses that can breach deep
thresholds which no individual source could reach alone. A single 1000
MW interconnector trip at average system conditions (\(H = 180\) GVA\(\cdot\)s,
\(D = 28\) GW, \(R = 1500\) MW) produces a median nadir of approximately
0.36 Hz, insufficient to breach the 49.2 Hz threshold. Two such trips occurring
simultaneously produce a 2000 MW effective loss with a median nadir of
approximately 0.73 Hz, directly at the SQSS threshold.

Each pair is classified by dependency type, which determines its
co-occurrence rate. \emph{Independent coincidence} pairs (the majority)
represent random temporal overlap with no physical coupling, at rates of
0.01--0.10/yr depending on severity. \emph{Common-cause} pairs share
infrastructure: the IFA bipole pair (BP1 + BP2) sharing the Channel
Tunnel route and converter stations is the canonical example, with
rates up to 0.25/yr for moderate events. \emph{Proximity} pairs
represent geographically close sources susceptible to correlated
weather-driven failures. \emph{Operator-coupled} pairs share a common
fleet operator whose operational decisions may correlate trip timing.

Severity is further differentiated into three classes: moderate
(combined loss 1000--1500 MW, rates 0.10--0.25/yr), severe (1500--2000
MW, 0.03/yr), and extreme (\textgreater2000 MW, 0.01/yr).
The co-occurrence rates for independent pairs are derived from the product of individual source rates scaled by the settlement-period overlap probability; for common-cause pairs (e.g., IFA bipoles sharing Channel Tunnel infrastructure), rates are informed by historical common-mode failure data. Proximity and operator-coupled rates are set at intermediate values between independent and common-cause based on the assessed degree of physical or operational coupling. Sensitivity analysis confirms that the exact choice of Layer~B rates has limited impact on the 49.2~Hz result (Layer~B contributes primarily at 48.8~Hz where combined losses exceed 2000~MW).
Combined-loss PMFs are computed by discrete convolution of the two
constituent sources' individual PMFs, preserving the empirical output
profiles rather than assuming additive capacities.

Layer B pairs become the dominant contributors to the 48.8 Hz hazard: at
this deep threshold, the five extreme pairs (with combined mean losses
exceeding 2000 MW) collectively account for a larger fraction of the
exceedance rate than any individual Layer A source. This is consistent with the physical expectation 
that very deep frequency deviations require either an extremely
large single loss or simultaneous failures.

\emph{2) Layer C, Compound Cascade:} Compound cascade events model
the mechanism observed on 9 August 2019, in which a large initial loss
produces high RoCoF that triggers distributed energy resource (DER)
protection relays, disconnecting an additional 200--500 MW of embedded
generation. The cascade is gated by a RoCoF threshold of 0.125 Hz/s,
which reflects the typical RoCoF protection relay setting on GB-connected
DER installations:

\[P(\text{cascade} \mid \Delta P, H) = \begin{cases} 0 & \text{if RoCoF} < 0.125 \text{ Hz/s} \\ p_{\text{cond}} & \text{if RoCoF} \geq 0.125 \text{ Hz/s} \end{cases} \tag{5}\]

where RoCoF \(= \Delta P \cdot f_0 / (2H \times 1000)\) and
\(p_{\text{cond}}\) is
the conditional cascade probability. At \(H = 150\) GVA\(\cdot\)s, a loss of
approximately 750 MW is required to trigger the RoCoF threshold;
events below this size cannot initiate cascade regardless of other
conditions. Layer C is retained as a sensitivity analysis rather than
a logic tree branch, as \(p_{\text{cond}}\) and the additional
DER-loss magnitude remain poorly constrained by the available observations.

\section{IV. Frequency Response Prediction
Equation}\label{iv.-frequency-response-prediction-equation}

\subsection{A. The FRPE as the GMPE
Analogue}\label{a.-the-frpe-as-the-gmpe-analogue}

In PSHA, the GMPE predicts ground-motion intensity given earthquake
magnitude and source-to-site distance. In PFHA, the Frequency Response
Prediction Equation (FRPE) predicts the median frequency nadir
\(\mu(\Delta P, \mathbf{s})\) given disturbance size \(\Delta P\) and
system state \(\mathbf{s}\). Following PSHA practice, the conditional
exceedance probability is computed under a log-normal model as in (2).
Two FRPE implementations are used in the logic tree: an
analytical SFR model (40\% weight) and a physics-based lookup model
(60\% weight).

\subsection{B. Analytical SFR Model}\label{b.-analytical-sfr-model}

The analytical SFR model is a closed-form approximation based on the
Anderson-Mirheydar simplified frequency response model \cite{anderson_mirheydar_1990}. The
swing equation governing frequency dynamics in a synchronous system is:

\[M \cdot \frac{d(\Delta f)}{dt} = \Delta P - D_{\mathrm{eff}} \cdot \Delta f \tag{6}\]

where \(M = 2H \times 1000 / f_0\) is the effective rotational inertia
(with \(H\) the system kinetic energy in GVA\(\cdot\)s and
\(f_0 = 50\) Hz) and
\(D_{\mathrm{eff}}\) is the total effective damping:

\[D_{\mathrm{eff}} = \frac{d_{\mathrm{load}}}{100} \cdot P_D + \frac{P_R}{\mathrm{droop} \cdot f_0} \tag{7}\]

comprising load damping (\(d_{\mathrm{load}} = 1.0\)\% per Hz, \(P_D\) =
demand in MW) and governor response gain (\(P_R\) = response holdings in
MW, droop \(= 0.04\)). The median nadir is:

\[\mu_{\mathrm{SFR}} = \frac{\Delta P}{D_{\mathrm{eff}}} \cdot \sqrt{1 + \left(\frac{\tau_R}{\tau_{\mathrm{sys}}}\right)^2} \cdot b \tag{8}\]

where \(\tau_R = 1.0\) s is the response delivery delay,
\(\tau_{\mathrm{sys}} = M / D_{\mathrm{eff}}\) is the system time
constant, and \(b\) is a bias correction factor.

The SFR systematically overpredicts nadir severity because it does not
model enhanced frequency-sensitive response, fast-acting Dynamic
Containment delivery profiles, or non-linear load damping. Replay of 283
GC0105 events (Section IV-D) reveals a magnitude-dependent bias:
residuals follow \(\epsilon = 3.224 - 0.684 \cdot \ln(\Delta P)\),
indicating that overprediction worsens for larger events. A flat
(magnitude-averaged) bias correction of \(b = 0.370\) is applied as a compromise, with the logic tree branch exploring
\(b \in \{0.30, 0.37, 0.50\}\).

\subsection{C. Physics-Based Lookup Model}\label{c.-physics-based-lookup-model}

The physics-based FRPE is a frequency dynamics simulator that models the full physics of
frequency response: governor delivery profiles, Dynamic Containment
injection with creep characteristics, Dynamic Moderation, Dynamic
Regulation, static response triggered at 49.7/49.6~Hz, and physics-based load
damping (proportional power--frequency coefficient of 0.025). It therefore does not require a bias correction. Table~\ref{tab:sfr-vs-ffse} summarises the physical processes modelled by each FRPE.

\begin{table}[t]
\centering
\caption{Physical processes modelled by each FRPE implementation.}
\label{tab:sfr-vs-ffse}
\begin{tabular}{lcc}
\toprule
Physical Process & SFR & Physics-based \\
\midrule
Governor response & Single droop & Full delivery profile \\
DC delivery & Not modelled & Explicit with creep \\
Dynamic Moderation & Not modelled & Included \\
Dynamic Regulation & Not modelled & Included \\
Static response & Not modelled & At 49.7/49.6~Hz \\
Load damping & Linear 1\%/Hz & Physics-based \\
Freq.-sensitive demand & Not modelled & Included \\
\midrule
Bias correction needed & Yes ($b = 0.370$) & No \\
Logic tree weight & 40\% & 60\% \\
\bottomrule
\end{tabular}
\end{table}

A dense 5-dimensional grid of nadir values is pre-computed across the operationally relevant parameter space:

\[\underbrace{\text{loss}}_{\text{7 values}} \times \underbrace{\text{inertia}}_{\text{7}} \times \underbrace{\text{demand}}_{\text{5}} \times \underbrace{\text{response}}_{\text{5}} \times \underbrace{\text{DC}}_{\text{5}} = 6{,}125 \text{ primary} + 7{,}525 \text{ boundary} = 13{,}650 \text{ points}\]

covering loss magnitudes 200--1800~MW, inertia 80--350~GVA$\cdot$s, demand 15--45~GW, response 500--3000~MW, and DC 0--1200~MW. At runtime, 5-dimensional linear interpolation provides nadir estimates at any point within the grid, with missing boundary cells filled via nearest-neighbour extrapolation.

The integration of Dynamic Containment (DC) into the physics-based pathway
requires careful architectural treatment to avoid double-counting. For
physics-based paths, DC enters as the fifth grid coordinate with effective DC
computed as contracted DC volume \(\times\) DC effectiveness
(e.g., 1000 MW \(\times\) 0.85 = 850 MW), and the control model's DC
contribution is disabled in those physics-based paths. For SFR paths, DC enters via
the control model's response augmentation with the FRPE receiving no DC
input. This routing ensures DC is counted exactly once regardless of the
FRPE model used on a given logic tree path.

\subsection{D. Aleatory Variability}\label{d.-aleatory-variability}

The aleatory variability \(\sigma\) represents irreducible
event-to-event scatter in nadir prediction --- the PSHA ``sigma'' ---
arising from unmodelled governor heterogeneity, demand composition,
frequency-sensitive load behaviour, and measurement uncertainty. It is
calibrated from the residuals of 283 GC0105 event replays:

\[\sigma(\Delta P, \mathbf{s}) = \sigma_0 \cdot \left[1 + 0.2 \cdot \max\left(0, \frac{150 - H}{150}\right)\right] \cdot \left[1 + 0.1 \cdot \max\left(0, \frac{\Delta P - 500}{1000}\right)\right] \tag{9}\]

where \(\sigma_0 = 0.296\) is the base aleatory variability calibrated from
the standard deviation of log-residuals after magnitude-bias correction.
Both SFR and physics-based paths use the same \(\sigma_0\), but the physics-based model applies
milder scaling: inertia coefficient of 0.1 (versus 0.2 for SFR) and no
size scaling term, reflecting the more complete physics modelling
which leaves less residual aleatory scatter. Inertia scaling captures
the physical expectation that prediction uncertainty increases at low
inertia where small modelling errors have larger effect on the nadir.
Size scaling (SFR only) reflects increasing non-linearity for large
disturbances that push the system far from equilibrium.

The event replay pipeline works as follows: for each GC0105 event with
recorded RoCoF, inertia, and post-event frequency, the loss magnitude is
derived from the swing equation
(\(\Delta P = |\mathrm{RoCoF}| \times 2H \times 1000 / f_0\)), the FRPE predicts the
median nadir, and the log-residual
\(\ln(\Delta f_{\mathrm{obs}} / \mu)\) is recorded. The resulting
residual distribution has standard deviation 0.296 after removing the
magnitude-dependent bias trend, confirming that the log-normal
assumption in (2) is reasonable for central tendencies, although
Shapiro-Wilk testing (\(p < 0.05\)) indicates heavier tails than the
normal model, a known limitation driven by data quality outliers.

\begin{figure}[t]
\centering
\includegraphics[width=0.82\textwidth]{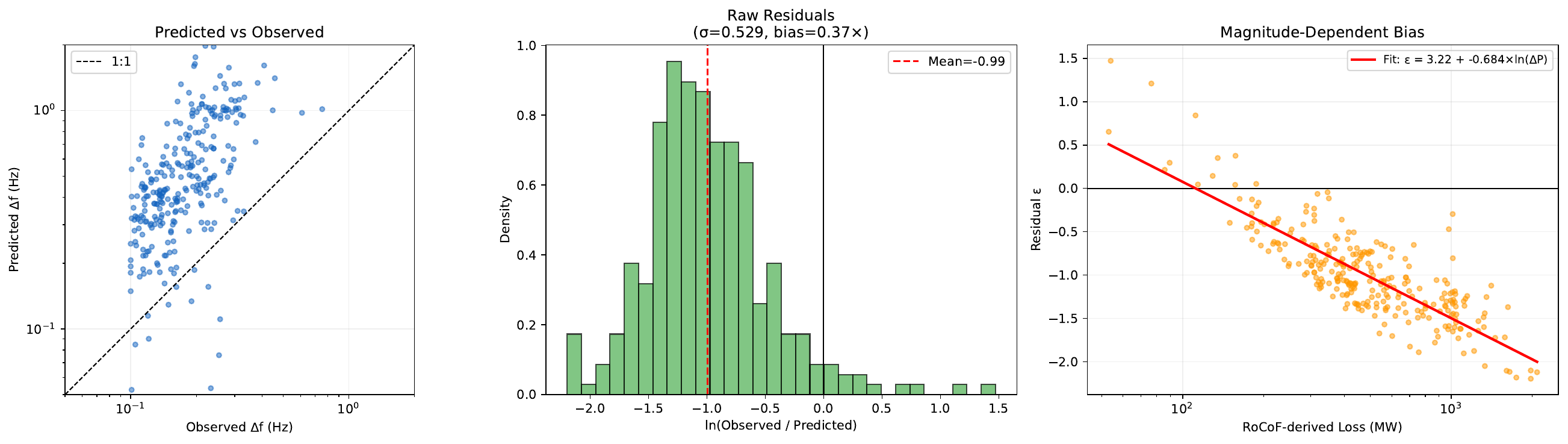}
\caption{Event-replay calibration of the analytical FRPE. The replay suite shows systematic overprediction of nadir severity by the uncorrected SFR model and provides the residual distribution used to calibrate the aleatory variability term.}
\label{fig:event-replay}
\end{figure}

\section{V. System State
Integration}\label{v.-system-state-integration}

\subsection{A. State Variables and Data
Sources}\label{a.-state-variables-and-data-sources}

The system state vector \(\mathbf{s} = (H, D, R)\) captures three
operating conditions that critically determine the frequency response to
a disturbance:

\begin{itemize}
\item
  \emph{Inertia H} (GVA\(\cdot\)s): the total synchronous kinetic energy,
  determining the initial RoCoF for a given loss. Sourced from NESO's
  half-hourly system inertia dataset, using market-position inertia (representing conditions
  at the time of commitment rather than real-time outturn). Observed range: 80--350 GVA\(\cdot\)s, 
  with a pronounced decline in minimum values from 2022 to 2026.
\item
  \emph{Demand D} (GW): national electricity demand, which affects both
  load damping (higher demand provides more MW/Hz of natural
  frequency-sensitive load reduction) and LFDD effectiveness (each
  percentage stage sheds more absolute MW at higher demand). Sourced
  from settlement-period national demand records. Observed range: 15--45
  GW with strong diurnal and seasonal patterns.
\item
  \emph{Frequency response R} (MW): total procured frequency response
  holdings including primary, secondary, and Dynamic Containment.
  Sourced from EAC auction results. Observed range: 500--3000 MW, with an increasing trend as
  the DC market has grown since 2020.
\end{itemize}

\subsection{B. Empirical Joint
Distribution}\label{b.-empirical-joint-distribution}

The joint distribution \(g(\mathbf{s})\) is estimated non-parametrically
from 71,476 half-hourly settlement periods spanning 2022--2026.
Quantile-based binning with \(N_{\mathrm{bin}} = 50\) bins is constructed by sorting
settlement periods on a weighted composite state metric and partitioning
into bins of approximately equal observation count. Each bin weight
\(w_k\) represents the fraction of operational time spent in that
regime.

This non-parametric approach preserves observed correlations between
state variables. Three correlations are particularly important for
frequency risk:

\begin{itemize}
\item
  \emph{Low inertia \(\leftrightarrow\) high wind penetration:} periods of high renewable
  output displace synchronous generation, reducing both inertia and the
  system's inherent resistance to frequency deviations. This correlation
  concentrates risk: the conditions that make frequency events more
  severe (low \(H\)) are the same conditions under which large wind
  farms (potential loss sources) are operating at high output.
\item
  \emph{Low inertia \(\leftrightarrow\) lower demand:} summer afternoons with high solar
  and wind output tend to coincide with lower national demand, reducing
  both inertia and the natural load-damping MW/Hz response.
\item
  \emph{Response holdings \(\leftrightarrow\) market conditions:} DC and DR procurement
  responds to anticipated system conditions, providing partial but
  imperfect correlation between response holdings and periods of higher
  risk.
\end{itemize}

\subsection{C. State-Dependent Effects on
Hazard}\label{c.-state-dependent-effects-on-hazard}

State integration is essential because the frequency nadir is a strongly
non-linear function of inertia. For a 1000 MW loss, the median nadir at
\(H = 120\) GVA\(\cdot\)s is approximately 1.5 times deeper than at \(H = 250\)
GVA\(\cdot\)s. Since the Gaussian CDF in the exceedance probability (2) is
non-linear, this means that the lowest-inertia settlement periods
contribute disproportionately to deep-threshold exceedance rates.

A computation using mean operating conditions (\(\bar{H} = 180\) GVA\(\cdot\)s,
\(\bar{D} = 28\) GW, \(\bar{R} = 1200\) MW) would systematically
underestimate the 49.2 Hz and 48.8 Hz hazard by failing to capture the
low-inertia tail where most severe risk concentrates. The state
integration in (1) correctly weights each operating regime by its
observed probability of occurrence, ensuring that rare but
high-consequence low-inertia periods contribute proportionally to the
hazard.

State disaggregation (Section VII-C) confirms this: at 49.2 Hz,
settlement periods with \(H < 140\) GVA\(\cdot\)s contribute over 60\% of the
total hazard despite representing less than 25\% of operational time.

\section{VI. Controls and Risk
Reduction}\label{vi.-controls-and-risk-reduction}

\subsection{A. Dynamic Containment}\label{a.-dynamic-containment}

Dynamic Containment (DC) is a fast-acting frequency response service
procured via the Enduring Auction Capability (EAC), with a contracted
volume of approximately 1000 MW \cite{neso_eac_2024}. DC delivery effectiveness,
the fraction of contracted volume that is actually delivered during an
event, is modelled as a logic tree branch with values
\(\{0.70, 0.85, 0.95\}\), where the central value of 0.85 is informed by
EAC data. The routing of DC through the SFR and physics-based
pathways is described in Section IV-C.

\subsection{B. Low-Frequency Demand
Disconnection}\label{b.-low-frequency-demand-disconnection}

The GB LFDD scheme disconnects demand automatically across nine frequency
stages from 48.8~Hz to 47.8~Hz, shedding up to 60\% of total demand.
The PFHA models a simplified five-stage representation aggregated at
48.8, 48.6, 48.4, 48.2, and 48.0~Hz, capturing the cumulative
demand reduction at each threshold. LFDD is modelled via two mechanisms: an
exceedance factor that reduces the probability of breaching a threshold
when preceding LFDD stages have activated, and a nadir cap that limits
the predicted median nadir when LFDD-shed demand reduces the effective
power imbalance.

The exceedance factor uses a strict inequality: each LFDD stage can only prevent breaches at thresholds deeper than its own activation frequency, since frequency must reach the activation level before the relay triggers. Stage 1 (48.8~Hz) therefore protects 48.6~Hz and below but cannot prevent the 48.8~Hz breach itself. This treatment is consistent with the 9 August 2019 event, in which LFDD Stage 1 triggered and the 48.8~Hz threshold was breached simultaneously. Relay effectiveness is modelled as a logic tree branch with values \(\{0.70, 0.85, 0.95\}\).

\subsection{C. Risk-Reduction
Quantification}\label{c.-risk-reduction-quantification}

A distinctive capability of the PFHA framework is the quantification of
the marginal risk reduction from individual control measures. By running
the hazard integral (1) under four configurations (no controls, DC
only, LFDD only, and both DC and LFDD), the risk reduction
attributable to each measure is isolated.

Table~\ref{tab:defence-value} presents the risk-reduction decomposition at the three key
thresholds.

\begin{table}[t]
\centering
\caption{Risk-Reduction Decomposition (Central Logic Tree Parameters)}
\label{tab:defence-value}
\begin{tabular}{lccc}
\toprule
Configuration & 49.5 Hz & 49.2 Hz & 48.8 Hz \\
\midrule
No controls (uncontrolled) & 7.3/yr & 0.93/yr & 0.25/yr \\
DC only & 4.2/yr & 0.41/yr & 0.08/yr \\
LFDD only & 3.8/yr & 0.36/yr & 0.03/yr \\
DC + LFDD & 2.26/yr & 0.213/yr & 0.020/yr \\
\textbf{Combined risk reduction} & \textbf{69\%} & \textbf{77\%} & \textbf{92\%} \\
\bottomrule
\end{tabular}
\end{table}

The results reveal several consistent patterns with clear physical explanations. First, LFDD's
marginal value increases with severity: it provides modest reduction at
49.5 Hz (where no LFDD stages activate, but LFDD's nadir-capping effect
at deeper frequencies indirectly reduces the breach probability) and
dominant reduction at 48.8 Hz (where multiple relay stages progressively
disconnect demand, arresting the frequency decline). Second, DC provides
a more uniform but moderate contribution across thresholds
(approximately 1.4\(\times\) reduction at 49.2 Hz), consistent with its
role as a fast-acting containment service that limits the initial
frequency excursion regardless of severity. Third, there is a positive
interaction effect: the combined reduction exceeds the sum of the
individual marginal reductions, because DC limits the nadir depth and
thereby prevents frequency from reaching the deeper LFDD stages in some
scenarios.

This decomposed view provides quantitative evidence for procurement decisions.
Because the PFHA computes hazard as a function of control parameters, it can evaluate marginal changes in DC or LFDD policy. For example, it can answer
the question: ``If DC procurement were reduced from 1000 MW to 700 MW,
by how much would the 49.2 Hz exceedance rate increase?'', a question
that requires the hazard to be computed as a function of contracted DC volume,
which the PFHA framework supports natively through the DC effectiveness
logic tree branch.

\begin{figure}[t]
\centering
\includegraphics[width=0.88\textwidth]{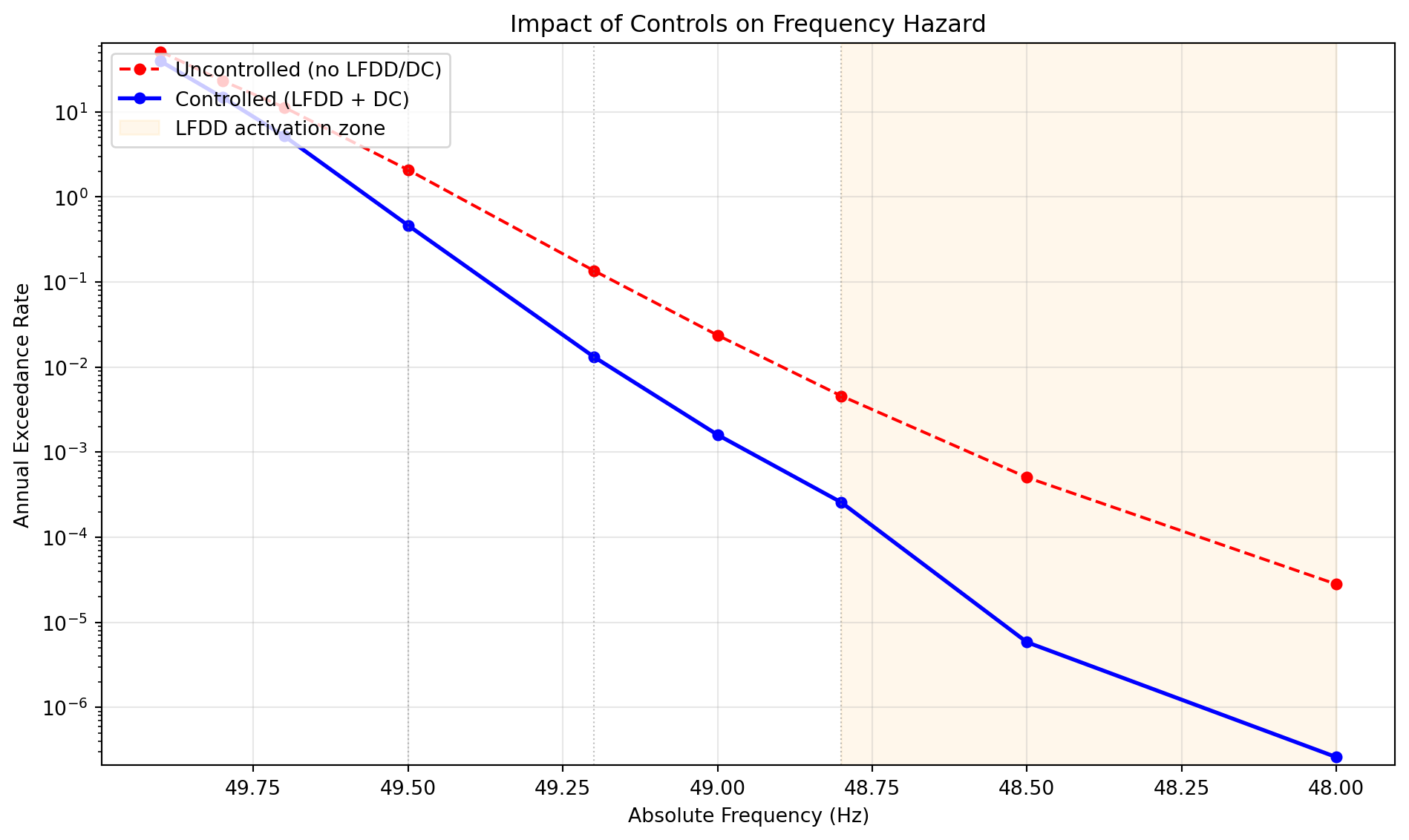}
\caption{Risk reduction from Dynamic Containment and LFDD controls. The controlled hazard curve lies materially below the uncontrolled curve, with the largest reduction occurring at the deepest thresholds where LFDD stages activate.}
\label{fig:defence-value-curves}
\end{figure}

\section{VII. Logic Tree and Uncertainty
Quantification}\label{vii.-logic-tree-and-uncertainty-quantification}

\subsection{A. Logic Tree Design}\label{a.-logic-tree-design}

The logic tree captures epistemic uncertainty through six branching
levels, producing
\(2 \times 3 \times 3 \times 2 \times 3 \times 3 = 324\) paths, each
yielding a distinct hazard curve. Table~\ref{tab:logic-tree} specifies the branches.

\begin{table}[t]
\centering
\caption{Logic Tree Specification}
\label{tab:logic-tree}
\begin{tabular}{p{3.5cm}p{3cm}p{6cm}}
\toprule
Branch & Parameter & Options (weights) \\
\midrule
1. FRPE model & Model type & SFR (0.40), Physics-based (0.60) \\
2. Aleatory sigma & $\sigma_0$ & 0.20 (0.25), 0.296 (0.50), 0.40 (0.25) \\
3. Bias correction & $b$ & 0.30 (0.30), 0.37 (0.40), 0.50 (0.30) \\
4. Occurrence model & Type & Poisson (0.70), Compound (0.30) \\
5. DC effectiveness & Fraction & 0.70 (0.25), 0.85 (0.50), 0.95 (0.25) \\
6. LFDD effectiveness & Fraction & 0.70 (0.25), 0.85 (0.50), 0.95 (0.25) \\
\bottomrule
\end{tabular}
\end{table}

Branch weights are assigned based on data support where available (sigma
central value from event replay, DC effectiveness from EAC data) and
on the assessed range of credible values where direct calibration data
is limited (occurrence model type, LFDD relay effectiveness). The physics-based model receives majority weight (0.60)
because it models the relevant physics directly without requiring bias
correction; the SFR is retained at 0.40 as an independent analytical
check. Branch 3 (bias correction) affects only SFR paths; physics-based paths
ignore this parameter entirely, which has significant implications for
the sensitivity analysis (Section VII-D).

\begin{figure}[t]
\centering
\begin{tikzpicture}[
  >=Latex,
  font=\scriptsize,
  lab/.style={align=center, font=\bfseries\footnotesize},
  opt/.style={draw, rounded corners, minimum width=1.65cm, minimum height=0.55cm, align=center},
  branch/.style={-{Latex[length=1.5mm]}, thick, gray!65},
  faint/.style={branch, opacity=0.25}
]
\node[lab] (l1) at (0,1.65) {FRPE\\model};
\node[lab] (l2) at (2.5,1.65) {Aleatory\\sigma};
\node[lab] (l3) at (5.0,1.65) {Bias\\correction};
\node[lab] (l4) at (7.5,1.65) {Occurrence\\model};
\node[lab] (l5) at (10.0,1.65) {DC\\effectiveness};
\node[lab] (l6) at (12.5,1.65) {LFDD\\effectiveness};

\node[opt, fill=green!8, draw=green!50!black] (a1) at (0,0.55) {SFR\\0.40};
\node[opt] (a2) at (0,-0.35) {Physics-based\\0.60};

\node[opt, fill=green!8, draw=green!50!black] (b2) at (2.5,0.10) {$\sigma=0.296$\\0.50};
\node[opt] (b1) at (2.5,0.95) {$\sigma=0.20$\\0.25};
\node[opt] (b3) at (2.5,-0.75) {$\sigma=0.40$\\0.25};

\node[opt, fill=green!8, draw=green!50!black] (c2) at (5.0,0.10) {$b=0.37$\\0.40};
\node[opt] (c1) at (5.0,0.95) {$b=0.30$\\0.30};
\node[opt] (c3) at (5.0,-0.75) {$b=0.50$\\0.30};

\node[opt, fill=green!8, draw=green!50!black] (d1) at (7.5,0.55) {Poisson\\0.70};
\node[opt] (d2) at (7.5,-0.35) {Compound\\0.30};

\node[opt] (e1) at (10.0,0.95) {0.70\\0.25};
\node[opt, fill=green!8, draw=green!50!black] (e2) at (10.0,0.10) {0.85\\0.50};
\node[opt] (e3) at (10.0,-0.75) {0.95\\0.25};

\node[opt] (f1) at (12.5,0.95) {0.70\\0.25};
\node[opt, fill=green!8, draw=green!50!black] (f2) at (12.5,0.10) {0.85\\0.50};
\node[opt] (f3) at (12.5,-0.75) {0.95\\0.25};

\node[draw, rounded corners, fill=purple!8, align=center, minimum width=2.2cm, minimum height=1.0cm] (out) at (15.2,0.10) {324 logic-tree\\hazard paths};

\foreach \n in {b1,b2,b3} {\draw[faint] (a1.east) -- (\n.west);}
\draw[branch, green!55!black] (a1.east) -- (b2.west);
\draw[branch, green!55!black] (b2.east) -- (c2.west);
\draw[branch, green!55!black] (c2.east) -- (d1.west);
\draw[branch, green!55!black] (d1.east) -- (e2.west);
\draw[branch, green!55!black] (e2.east) -- (f2.west);
\draw[branch, green!55!black] (f2.east) -- (out.west);
\end{tikzpicture}
\caption{Logic-tree structure. One highlighted path shows how FRPE choice, aleatory variability, bias correction, occurrence model, and control effectiveness propagate to the hazard output.}
\label{fig:logic-tree}
\end{figure}
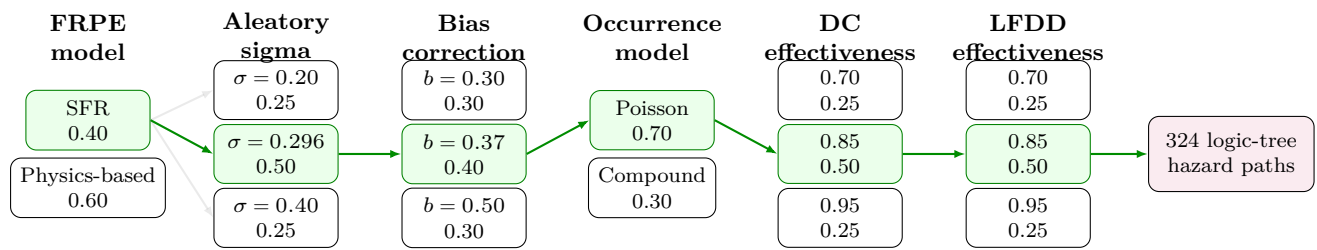

\subsection{B. Weighted Fractile
Computation}\label{b.-weighted-fractile-computation}

Each of the 324 paths produces a rate at each threshold. The weighted
distribution of these rates is characterised by percentiles computed
from the cumulative weight function: rates are sorted and cumulative
weights are summed to identify the \(p\)-th percentile. The mean rate
\(\bar{\lambda} = \sum_p w_p \lambda_p\) is the PSHA regulatory standard
(it accounts for fat tails in epistemic uncertainty), while the median
(50th percentile) is less sensitive to extreme paths.

The 90\% epistemic band (5th to 95th percentile) spans 19\(\times\) at
49.5 Hz, 53\(\times\) at 49.2 Hz, and 238\(\times\) at 48.8 Hz. This
wide band is driven primarily by the bias correction branch on SFR
paths: high-bias values produce very large rates, pulling the upper
tail. Critically, the 60\% of paths using the physics-based model do not carry bias
uncertainty, which narrows the effective uncertainty for
decision-making. The median is therefore a more stable and
decision-useful metric than the mean for stakeholder communication.

\subsection{C. Disaggregation}\label{c.-disaggregation}

Disaggregation decomposes the total hazard rate at a given threshold
into contributions from individual sources, loss magnitudes, system
states, and epsilon (the number of aleatory standard deviations above
the median prediction). This is the frequency-domain analogue of the
PSHA magnitude-distance-epsilon (\(M\)-\(R\)-\(\varepsilon\))
disaggregation that identifies the dominant earthquake scenarios
\cite{baker_bradley_stafford_2021}, \cite{baker_2013_psha}.

The PFHA implements disaggregation across five dimensions:

\emph{1) By source:} The fractional contribution
\(\lambda_i / \lambda_{\text{total}}\) identifies which sources drive
the hazard at each threshold. At 49.5 Hz, many sources contribute
because moderate events (400--800 MW) are sufficient to breach this
threshold at low inertia, and the high-rate CCGT fleet catch-all
dominates through sheer frequency of occurrence. At 49.2 Hz, the
dominant contributors shift to high-capacity sources: the IFA
interconnector (combined bipoles), the CCGT fleet aggregate, and the
nuclear stations --- all sources capable of producing losses exceeding
1000 MW. At 48.8 Hz, Layer B extreme pairs (combined losses exceeding
2000 MW) become significant contributors, reflecting the physical
reality that very deep deviations require either exceptionally large
single losses or simultaneous failures.

\emph{2) By loss size:} The dominant loss magnitude shifts
systematically with threshold severity. At 49.5 Hz, moderate events
(600--1000 MW) dominate because they are relatively common and can
breach this mild threshold with modest aleatory amplification. At 48.8
Hz, only very large events (\textgreater1200 MW) contribute meaningfully
--- smaller losses cannot reach this deep threshold even at maximum
aleatory excursion (\(+3\sigma\)).

\emph{3) By system state:} Low-inertia settlement periods (\(H < 140\)
GVA\(\cdot\)s) contribute over 60\% of the 49.2 Hz hazard despite representing
less than 25\% of operational time. This concentration intensifies at
deeper thresholds: at 48.8 Hz, the lowest-inertia quartile contributes
over 80\% of the rate. This result has direct operational implications
--- frequency risk is overwhelmingly concentrated in specific,
identifiable operating regimes.

\emph{4) By epsilon:} The epsilon disaggregation reveals how many
standard deviations above the median prediction the dominant scenarios
require. At 49.2 Hz, most hazard comes from
\(\varepsilon \in [0.5, 2.0]\) --- events that are 0.5--2 sigma worse
than the median prediction. Events requiring \(\varepsilon > 2.5\)
contribute minimally despite their extremity, because the Gaussian tail
probability falls faster than the exceedance probability rises. This
confirms that the hazard is not dominated by implausible tail events but
by scenarios that are moderately worse than expected, a practically
important finding for risk communication.

\emph{5) Combined size-inertia-epsilon:} The full 3D disaggregation
identifies the complete scenario triplet that dominates each threshold.
At 49.2 Hz, the modal scenario is approximately ``1200 MW loss at
\(H = 130\) GVA\(\cdot\)s with \(\varepsilon = +1.2\)'', a large
interconnector or nuclear trip during a low-inertia period that produces
a nadir about one sigma deeper than the median prediction. This level of
scenario specificity enables targeted risk management: the dominant risk
scenario is identifiable, and the system conditions under which it
occurs are operationally observable.

\begin{figure}[t]
\centering
\includegraphics[width=0.88\textwidth]{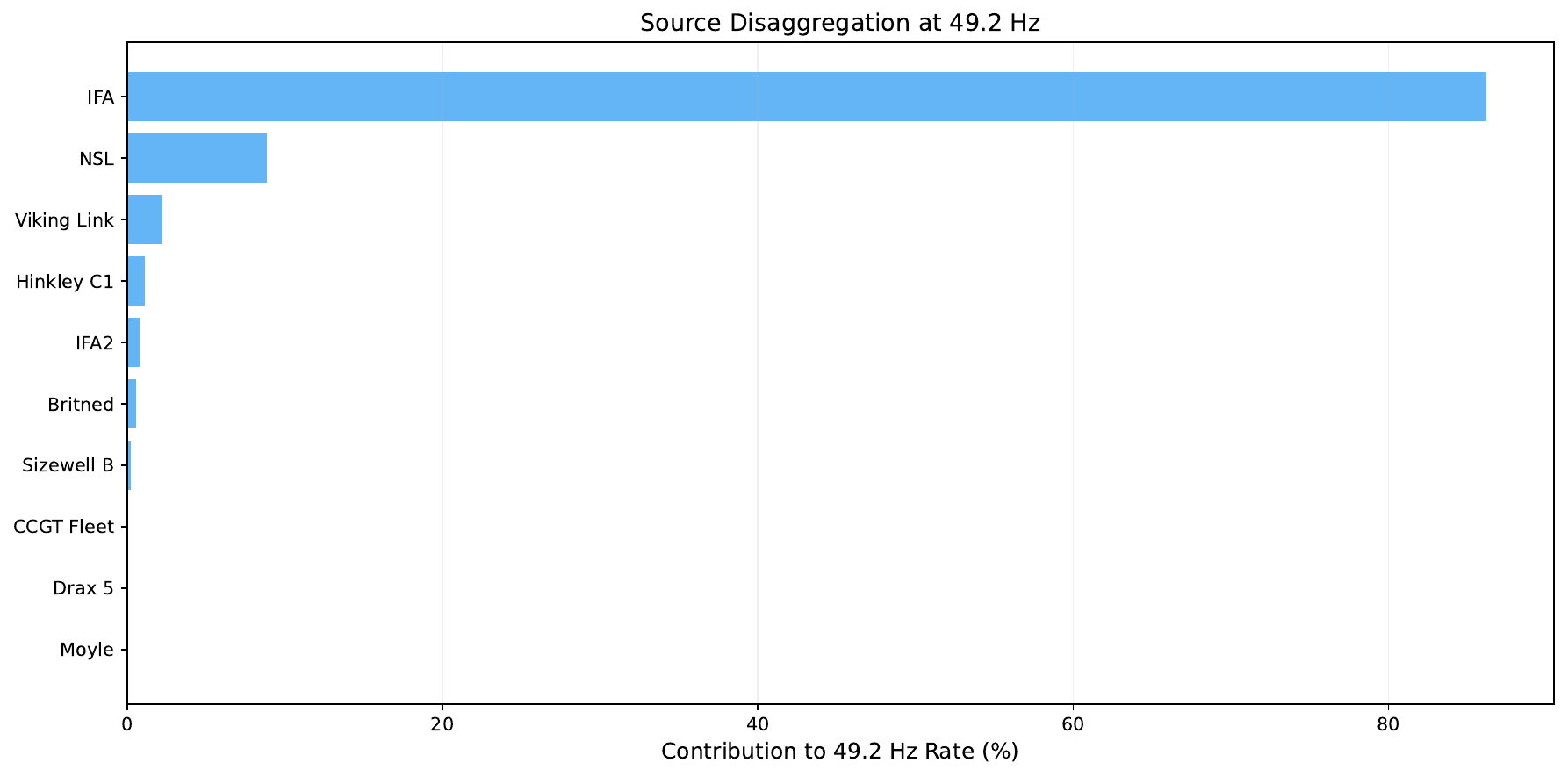}
\caption{Source disaggregation at the 49.2~Hz threshold. Interconnectors dominate the rate contribution, led by IFA, with secondary contributions from NSL, Viking Link, and large conventional units.}
\label{fig:source-disagg}
\end{figure}

\begin{figure*}[t]
\centering
\includegraphics[width=0.95\textwidth]{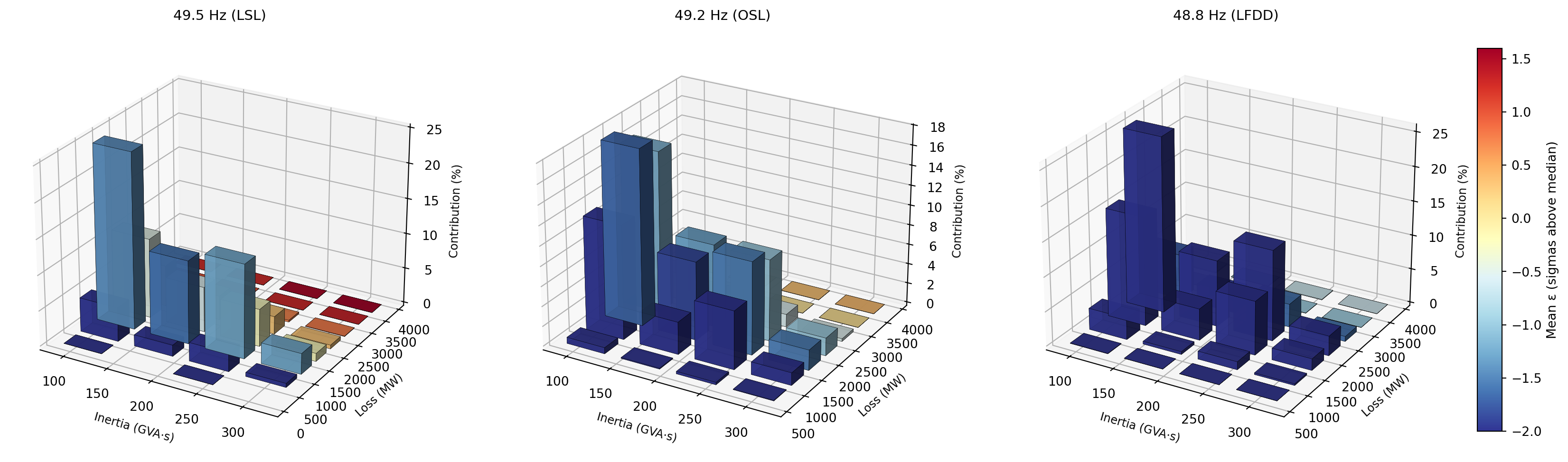}
\caption{PSHA-style 3D disaggregation at three regulatory thresholds. Each bar represents a (loss~size~$\times$~inertia) cell; height is fractional contribution to the exceedance rate; colour encodes mean $\varepsilon$. At 49.5~Hz (left), hazard is broadly distributed across many moderate cells. At 48.8~Hz (right), hazard concentrates sharply in large-loss, low-inertia cells at high $\varepsilon$.}
\label{fig:3d-disagg}
\end{figure*}

Fig.~\ref{fig:3d-disagg} presents the full PSHA-style 3D disaggregation at all three regulatory thresholds. Each bar represents a (loss~size, inertia) cell, with height proportional to its fractional contribution to the exceedance rate and colour encoding the mean epsilon (number of aleatory standard deviations above the median prediction). The shift from a broadly distributed hazard at 49.5~Hz (many moderate cells contribute) to a sharply concentrated hazard at 48.8~Hz (dominated by a few large-loss, low-inertia cells at high epsilon) shows that different scenario families dominate at different hazard levels. This visualisation provides system operators with an actionable risk portrait: at 49.2~Hz, the dominant risk comes from 1000--1500~MW losses occurring when system inertia is below 150~GVA$\cdot$s, with nadir outcomes approximately 1--1.5 standard deviations worse than the median prediction. These are operationally observable conditions against which targeted mitigation (increased DC procurement, minimum inertia constraints) can be directed.

\begin{figure}[t]
\centering
\includegraphics[width=0.82\textwidth]{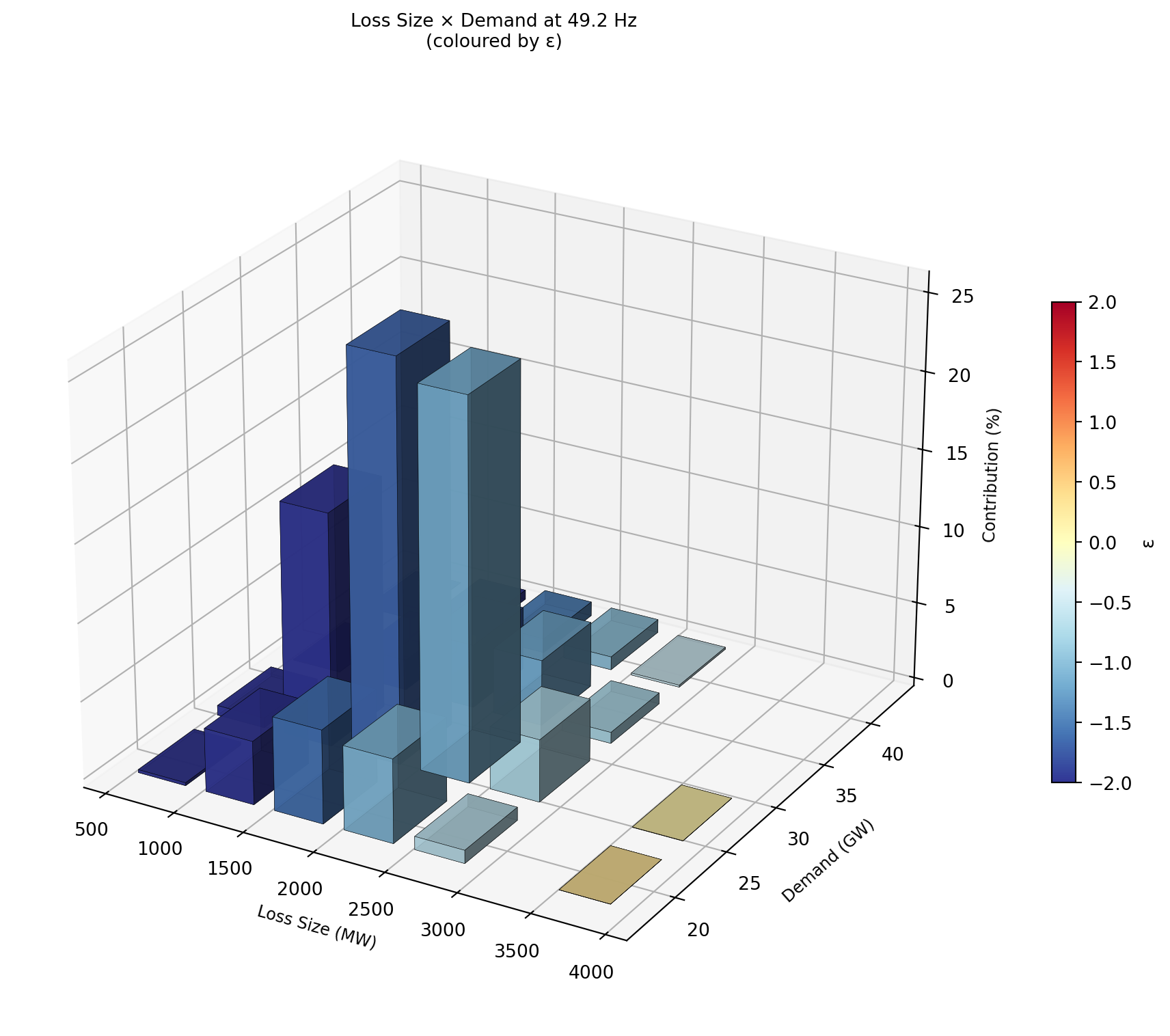}
\caption{3D disaggregation at 49.2~Hz: loss size versus demand, coloured by~$\varepsilon$. Hazard concentrates at large losses (1000--2000~MW) during moderate demand periods (25--35~GW) where load damping is limited.}
\label{fig:3d-size-demand}
\end{figure}

Fig.~\ref{fig:3d-size-demand} provides a complementary view at 49.2~Hz, disaggregating by loss size and demand. The concentration at high loss magnitudes (1000--2000~MW) and moderate demand levels (25--35~GW) reflects the physical mechanism: moderate demand provides less load damping per Hz while still representing the most common operating conditions during which large sources are online.

\subsection{D. Sensitivity Analysis}\label{d.-sensitivity-analysis}

Tornado analysis at 49.2 Hz, varying one parameter at a time while
holding others at their central (highest-weight) values, yields the
following swing factors: sigma produces a 2.9\(\times\) swing (from
0.086 to 0.245/yr), FRPE model type produces 2.4\(\times\) (from 0.057
for SFR-only to 0.135 for physics-based-only), DC effectiveness produces
1.4\(\times\), while bias correction shows a 1.0\(\times\) swing (no
sensitivity) because the central path uses the physics-based model (60\% weight), which
ignores bias entirely. This result directs future modelling effort:
improving sigma calibration (e.g., by running the event replay against
physics-based predictions rather than SFR) is the most impactful improvement
available.

\begin{figure}[t]
\centering
\includegraphics[width=0.78\textwidth]{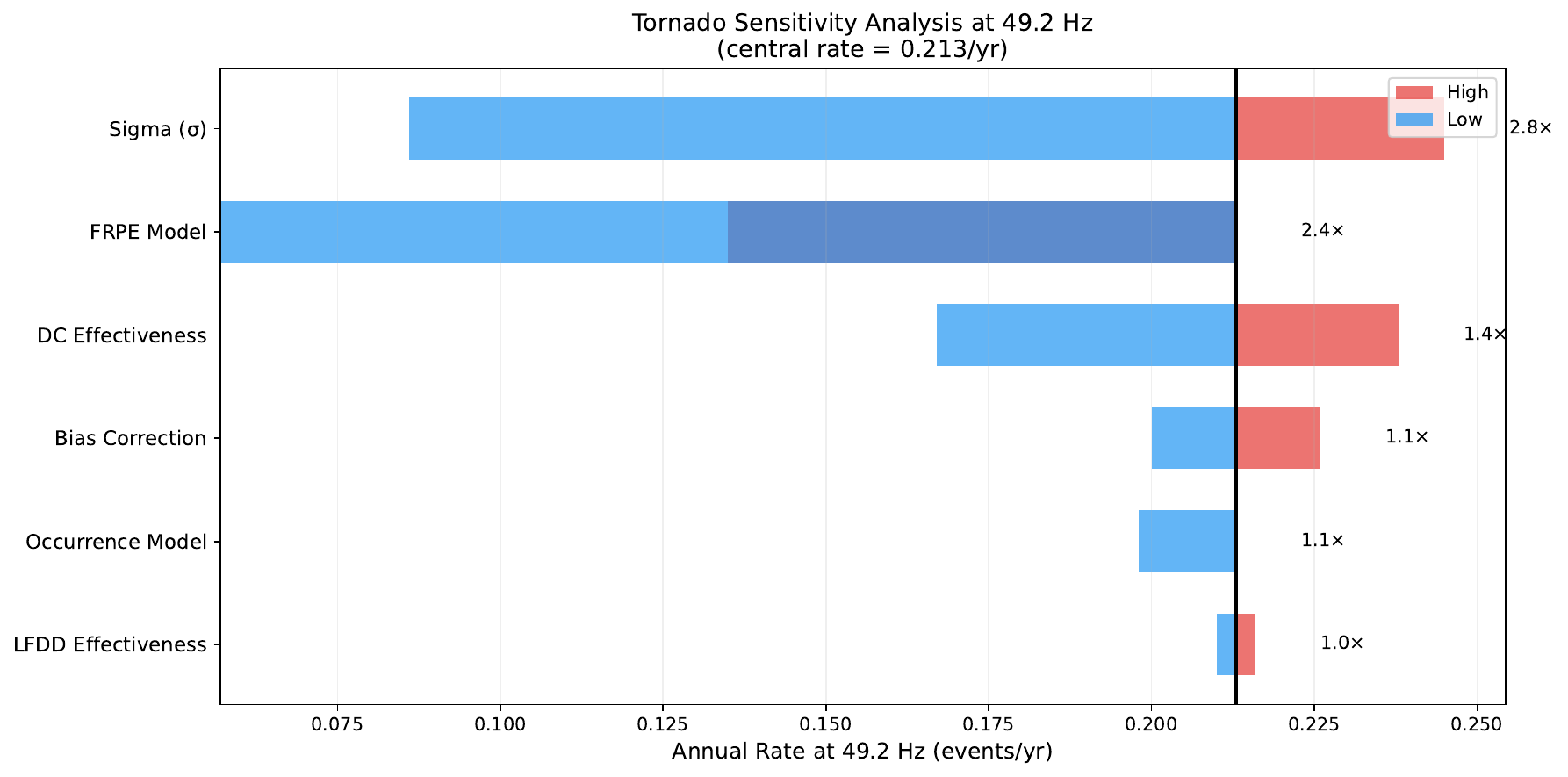}
\caption{Tornado sensitivity analysis at 49.2~Hz. Aleatory variability has the largest effect on the exceedance rate, followed by FRPE model choice and DC effectiveness.}
\label{fig:tornado}
\end{figure}

\section{VIII. Application to the GB Power
System}\label{viii.-application-to-the-gb-power-system}

\subsection{A. Results}\label{a.-results}

Table~\ref{tab:production-results} presents the PFHA results from the full 324-path
logic tree computation with the LFDD boundary condition correction
applied.

\begin{table}[t]
\centering
\caption{PFHA Results (324 Logic Tree Paths)}
\label{tab:production-results}
\begin{tabular}{lccccc}
\toprule
Threshold & Mean & Median & p05 & p95 & Return Period \\
\midrule
49.5 Hz & 2.26/yr & 2.68/yr & 0.52/yr & 9.8/yr & 0.44 yr \\
49.2 Hz & 0.213/yr & 0.168/yr & 0.023/yr & 1.22/yr & 4.7 yr \\
48.8 Hz & 0.020/yr & 0.009/yr & 0.001/yr & 0.238/yr & 50 yr \\
\bottomrule
\end{tabular}
\end{table}

The mean exceeds the median at 49.2 and 48.8 Hz because the rate
distribution across logic tree paths is right-skewed: a minority of
high-sigma, high-bias SFR paths produce very high rates that pull the
mean upward.

\begin{figure}[t]
\centering
\includegraphics[width=0.9\textwidth]{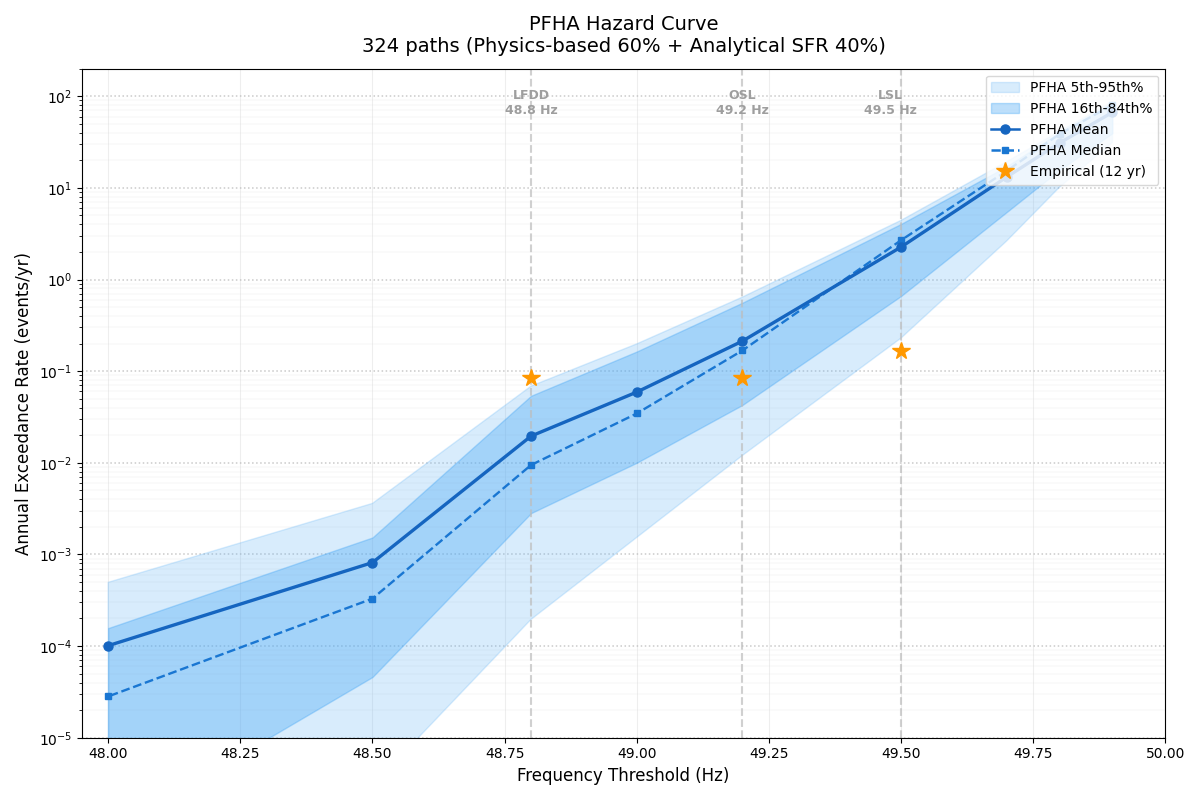}
\caption{PFHA hazard curve with epistemic uncertainty. The corrected LFDD boundary condition (strict inequality) appropriately increases the 48.8~Hz exceedance rate, consistent with the physical constraint that LFDD Stage~1 cannot prevent breach at its own activation frequency.}
\label{fig:production-hazard}
\end{figure}

\subsection{B. Cross-Validation with
FRCR}\label{b.-cross-validation-with-frcr}

Table~\ref{tab:pfha-vs-frcr} compares the PFHA results with the FRCR estimates.
The two methodologies use independently developed data-processing and
modelling pipelines: different physical models (PFHA uses the hazard
integral with FRPE; FRCR uses deterministic settlement-period enumeration), different
uncertainty treatments, and different software implementations.

\begin{table}[t]
\centering
\caption{PFHA vs FRCR Cross-Validation}
\label{tab:pfha-vs-frcr}
\begin{tabular}{lccc}
\toprule
Threshold & PFHA (mean) & FRCR & Ratio \\
\midrule
49.5 Hz & 2.26/yr & 2.85/yr & 0.79 \\
49.2 Hz & 0.213/yr & 0.138/yr & 1.54 \\
48.8 Hz & 0.020/yr & 0.039/yr & 0.51 \\
\bottomrule
\end{tabular}
\end{table}

The agreement at 49.2 Hz (within a factor of 1.54) is notable given the
complete independence of the two approaches. At 48.8 Hz, the PFHA
estimate is approximately half the FRCR value. This difference is partly
attributable to the structural difference in consequence modelling: the
FRCR uses a binary breach comparison where any event exceeding the
secured loss limit counts fully, while the PFHA assigns a continuous
exceedance probability that decreases as events approach the threshold
from above. At deep thresholds where many events are near the boundary,
this distinction becomes material. The FRCR estimate falls within the
PFHA's 90\% epistemic band at all thresholds.

\begin{figure}[t]
\centering
\includegraphics[width=0.86\textwidth]{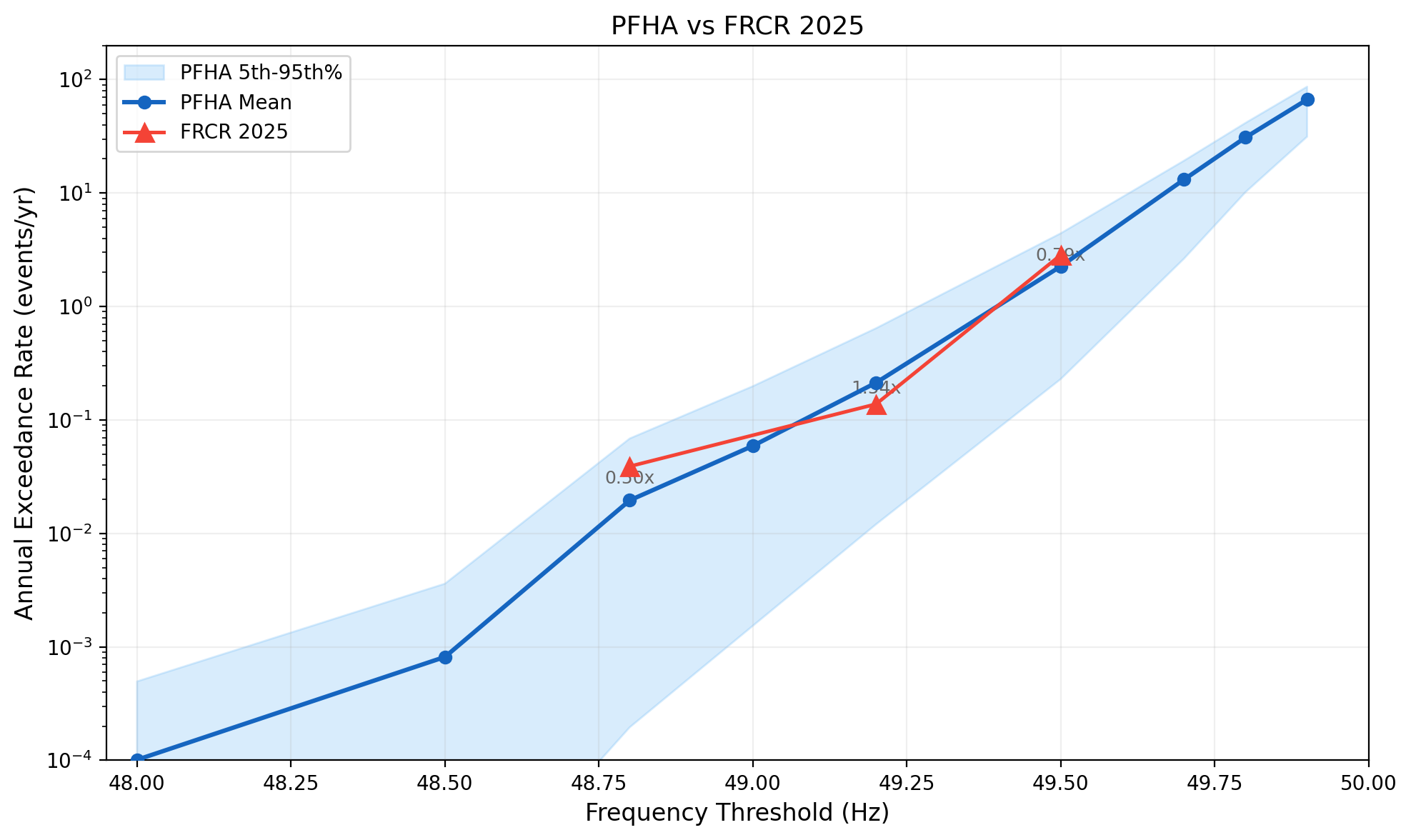}
\caption{Cross-validation between PFHA and FRCR across the decision thresholds. Despite using distinct data pipelines and model structures, the two approaches agree to within the same order of magnitude at all three thresholds and within a factor of~1.54 at 49.2~Hz.}
\label{fig:pfha-vs-frcr}
\end{figure}

\subsection{C. Risk-Reduction Results}\label{c.-risk-reduction-results}

The controlled versus uncontrolled comparison quantifies the risk
reduction provided by DC and LFDD, as detailed in Table~\ref{tab:defence-value} (Section
VI-C). At 49.2 Hz, the combined controls reduce the exceedance rate by
77\%. At 48.8 Hz, the reduction is 92\%, with LFDD providing the
majority of the benefit at this deep threshold (multiple relay stages
activate). DC contributes a more modest but uniform reduction across
thresholds (approximately 1.4\(\times\) at 49.2 Hz). The ability to decompose
risk reduction by individual control measure provides quantitative evidence
for DC procurement decisions and LFDD relay policy, complementing existing operational risk assessments.

\section{IX. Validation}\label{ix.-validation}

\subsection{A. Empirical Frequency
Scan}\label{a.-empirical-frequency-scan}

The most direct validation is comparison with observed threshold
breaches in the historical frequency record. A scan of 145 monthly files
of 1-second frequency data spanning 2014--2026 (approximately 12.1 years,
378 million samples) computes empirical exceedance rates at 30 thresholds
from 49.95 to 48.50~Hz in 0.05~Hz steps, using a 60-second event merge window.

The resulting empirical exceedance curve reveals a steep drop from approximately
14,200 events/yr at 49.95~Hz (normal operational frequency excursions) through
1.24/yr at 49.60~Hz (15 discrete events) to 0.167/yr at 49.50~Hz (2 events: the
9~August 2019 compound loss and the 22~December 2023 IFA bipole trip). All
thresholds from 49.20 to 48.80~Hz show a single event at 0.083/yr. Only the
9~August 2019 event reached those depths, as the December 2023 event recovered
at 49.275~Hz. The sharp cliff between 49.60 and 49.50~Hz marks the
boundary between routine frequency management and significant frequency events.

At 49.5~Hz, the empirical rate over 2014--2026 is 0.167/yr, materially
below both PFHA and FRCR. This indicates either conservatism in the
modelled shallow-threshold risk, non-stationarity between the historical
record and the 2022--2026 calibration window, or residual
mismatch between observed event counting and modelled exceedance
definitions.

At 49.2~Hz, the PFHA mean of 0.213/yr is 2.6 times higher than the
empirical 0.083/yr but falls within the Poisson 95\% confidence interval
for one event in 12 years ([0.002, 0.25]/yr). The PFHA rates are
expected to be higher than the 12-year average because the PFHA models
current (2022--2026) system conditions with lower inertia and
operational DC, while the empirical data averages over a period during
which inertia was generally higher and DC did not exist for most years.

The deep-threshold comparison is therefore best interpreted as a
consistency check rather than a calibration target. With \(n = 1\)
observed events below 49.2~Hz and 48.8~Hz, the empirical record cannot
tightly constrain the model, but it does confirm that such deep
excursions are rare and physically plausible.

\subsection{B. Out-of-Sample Temporal
Split}\label{b.-out-of-sample-temporal-split}

To test for overfitting, the GC0105 incident data is split temporally:
events through 2024 serve as the training set (approximately 260 events,
2.99 years), and 2025 events serve as the test set (approximately 89
events, 0.99 years). The Bayesian rates and FRPE sigma are recalibrated
on the training set only, and the full PFHA is re-run using single-path
central parameter values. Table~\ref{tab:oos} compares the results.

\begin{table}[t]
\centering
\caption{Out-of-sample temporal stability (single-path central rates)}
\label{tab:oos}
\begin{tabular}{lcccc}
\toprule
Threshold & Full Period & Training Only & Ratio & Stable? \\
\midrule
49.5 Hz & 1.67/yr & 1.39/yr & 0.83 & Yes \\
49.2 Hz & 0.113/yr & 0.093/yr & 0.82 & Yes \\
48.8 Hz & 0.009/yr & 0.008/yr & 0.89 & Yes \\
\bottomrule
\end{tabular}
\end{table}

All ratios are within 20\%, indicating temporal stability: removing
25\% of the data changes the output by less than 20\%. The training-only
rates are systematically lower, which is expected since fewer observed
events produce lower Bayesian posterior rate estimates. Zero threshold
breaches were observed in the 2025 test period, consistent with Poisson
expectations (P(0 events $|$ $\lambda$=0.09, T=1~yr) = 0.91 at 49.2~Hz).

\begin{figure}[t]
\centering
\includegraphics[width=0.88\textwidth]{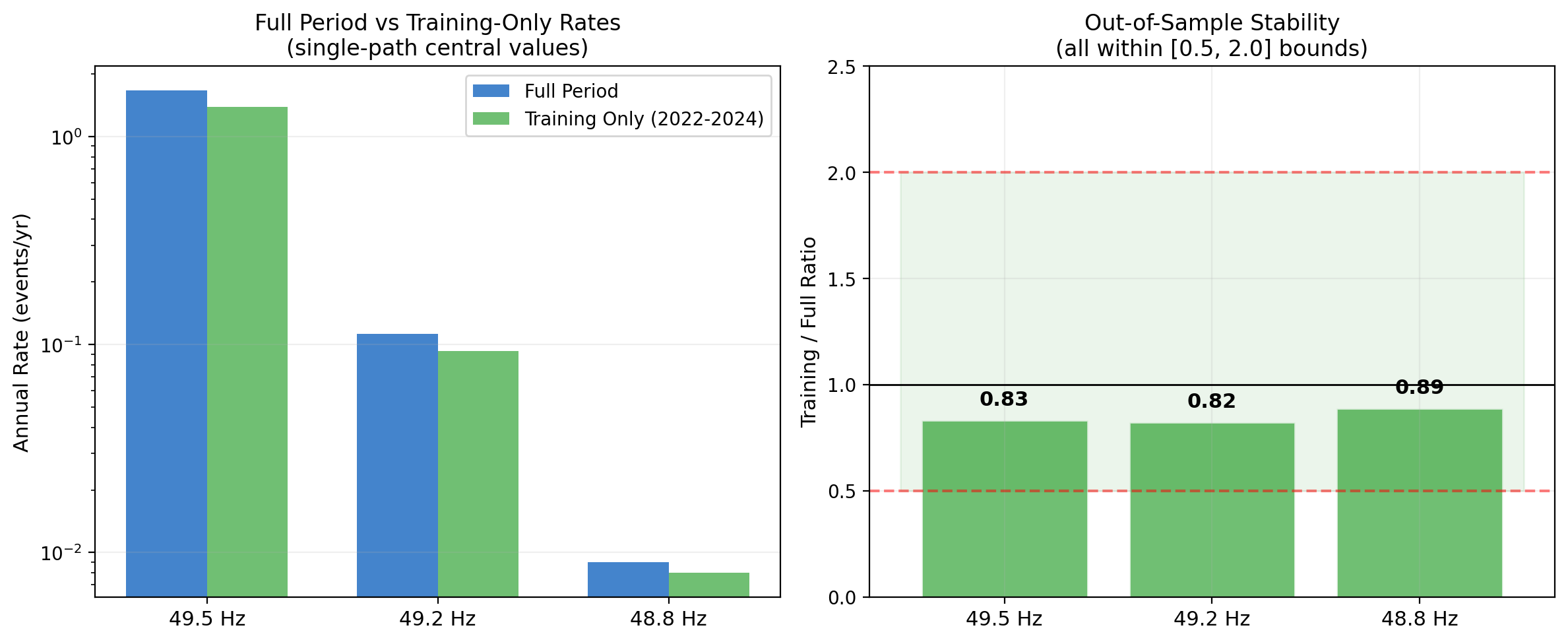}
\caption{Out-of-sample temporal stability. Left: training-only rates closely match full-period rates at all thresholds. Right: all training/full ratios fall within the [0.5, 2.0] stability bounds.}
\label{fig:oos-stability}
\end{figure}

\subsection{C. Physics-Based Model Event Replay}\label{c.-physics-based-event-replay}

\begin{figure}[t]
\centering
\includegraphics[width=0.88\textwidth]{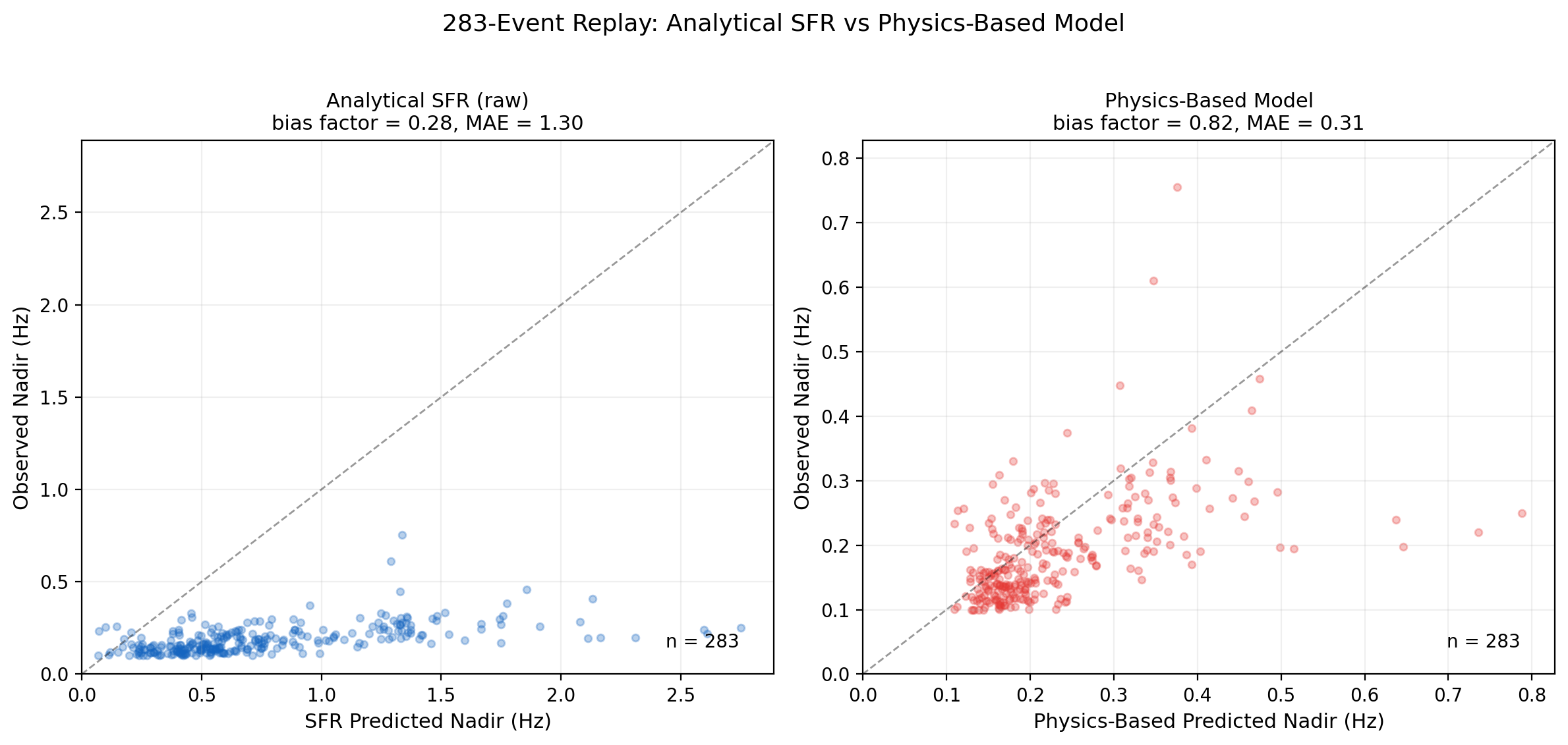}
\caption{283-event replay comparison. Left: the analytical SFR systematically overpredicts nadir severity (points below the 1:1 line). Right: the physics-based model clusters closer to the 1:1 line with a bias factor of 0.82 and 4.2$\times$ lower mean absolute error.}
\label{fig:frpe-comparison}
\end{figure}

The same 283 GC0105 events were replayed through the physics-based interpolator for comparison with the SFR. The physics-based model produces less biased predictions: mean log-residual $-0.20$ (bias factor 0.82) compared to the raw SFR's $-1.27$ (bias factor 0.28). The physics-based residual standard deviation is 0.33, compared to 0.53 for the raw SFR (or 0.296 after magnitude-bias correction). The mean absolute error is 0.31 for the physics-based model versus 1.30 for the raw SFR, a 4.2$\times$ improvement. These results justify the physics-based model's 60\% logic tree weight and its exemption from bias correction. Fig.~\ref{fig:frpe-comparison} shows the predicted-versus-observed scatter for both models.

\subsection{D. Actual MW Validation}\label{d.-actual-mw-validation}

For 74 of the 283 events, the actual MW loss was reconstructed by matching the event timestamp to B1610 settlement-period generation data (for generators) or InterconnectorFlows records (for interconnectors). This provides ground truth independent of the RoCoF-derived swing equation inversion. The magnitude-dependent bias slope steepens from $-0.684$ (RoCoF-derived) to $-1.018$ (actual MW), confirming that the bias is a real physical effect of the SFR's model limitations, not a measurement artefact of the RoCoF inversion. The corrected sigma increases slightly from 0.296 to 0.360 with actual MW, reflecting additional scatter from the B1610 matching process. This validation supports the use of flat bias correction ($b = 0.370$) rather than magnitude-dependent correction, which over-corrects at the hazard integration level.

\subsection{E. The 9 August 2019 Event as Physical Validation}\label{e.-9-aug-2019}

The 9 August 2019 compound loss event provides a unique physical validation case because it is the only event in the 12-year record to breach the 48.8~Hz LFDD threshold. The event sequence (Hornsea~1 wind farm, $\sim$900~MW, followed within 2~seconds by Little Barford CCGT, 737~MW, compounded by DER disconnection of $\sim$210~MW, at a system inertia of 210~GVA$\cdot$s) is precisely the type of compound cascade that the Layer~B (simultaneous pairs) and Layer~C (RoCoF-gated cascade) models are designed to capture.

The SFR model (with flat bias $b = 0.370$) predicts a median nadir of 0.64~Hz for the combined 1341~MW loss at the recorded system conditions, underestimating the observed 1.2~Hz deviation. This is consistent with the SFR's inability to model the staggered loss timing and DER cascade dynamics. The physics-based prediction for the same total loss is 0.41~Hz, further below the observed nadir because both FRPEs model an instantaneous single loss rather than the actual multi-second cascade sequence, and the physics-based model's DC treatment further reduces the predicted nadir.

The event also confirms the LFDD strict-inequality treatment: LFDD Stage~1 triggered \emph{at} 48.8~Hz but could not prevent the frequency from reaching that threshold, since it activated simultaneously with the breach. Frequency did not fall below 48.8~Hz, confirming that Stage~1 protects deeper thresholds (48.6~Hz and below) but cannot prevent breach at its own activation frequency.

\section{X. Discussion and
Conclusions}\label{x.-discussion-and-conclusions}

\subsection{A. Known Limitations}\label{a.-known-limitations}

Several limitations warrant acknowledgement. The 49.5~Hz comparison to
the historical empirical record remains materially high for both PFHA
and FRCR, indicating unresolved shallow-threshold conservatism or a
definition mismatch between modelled exceedance and observed event
counting. At 48.8~Hz, the PFHA estimate is approximately half the FRCR
value, partly attributable to the structural difference between
continuous exceedance probability (PFHA) and binary breach comparison
(FRCR) as discussed in Section~VIII-B. The log-normal assumption for
nadir prediction residuals is a reasonable central model but does not
capture the heavy tails evident in the Shapiro-Wilk test (\(p < 0.05\)),
driven partly by data quality outliers in the GC0105 dataset. LFDD is
modelled as an exceedance factor rather than a time-domain
relay-frequency race, a simplification that may over- or under-credit
LFDD depending on the event trajectory. Prior-dominated rates for
zero-trip sources (most of the 13 wind farms) rely entirely on
technology-class priors and could be significantly wrong if wind farm
trip behaviour differs from prior assumptions. Loss magnitudes for most
events are estimated from RoCoF-derived swing equation inversion rather
than directly measured MW, introducing systematic uncertainty. Finally,
the system state model assumes stationarity across the four-year
calibration period, which may not hold as the generation mix evolves.

\subsection{B. Future Work}\label{b.-future-work}

Priority extensions include: recalibration of sigma against physics-based
predictions (which would eliminate the need for bias correction entirely),
time-domain LFDD modelling to replace the exceedance factor approach with
simulation of the relay-frequency race, extension of B1610 matching to
provide actual MW for more events, and a richer cascade model where DER
disconnection volume is a function of RoCoF magnitude and system
composition. Projection of hazard curves under Future Energy Scenarios
(FES 2030, 2040) requires only updated state distributions, source
catalogues, and control parameters. Further extensions include per-source
rate branching using Bayesian credible intervals as logic tree branches
for top contributing sources, temporal non-stationarity modelling, and
application to other synchronous areas. The framework is applicable to
any system with operational generation, frequency, and incident data;
the EirGrid all-island system, Continental European synchronous area,
and Australian NEM are natural candidates, requiring adaptation of the
source catalogue, LFDD scheme, and state variable ranges but no changes
to the mathematical framework.

\subsection{C. Conclusions}\label{c.-conclusions}

This paper has presented Probabilistic Frequency Hazard Analysis (PFHA),
a framework that adapts the PSHA mathematical architecture
to power system frequency exceedance risk. The framework formulates the
hazard integral for frequency, implements it with a 51-source catalogue,
empirical loss distributions, Bayesian rate estimation, a dual
analytical and physics-based FRPE, and a 324-path logic tree, and
applies it to the GB power system.

The PFHA cross-validates with the independently-developed FRCR within a
factor of 1.5 at the 49.2 Hz threshold. It provides capabilities that
existing methods do not: continuous hazard curves, formal epistemic and
aleatory uncertainty quantification, source-level disaggregation, and
decomposed risk-reduction quantification.

The framework is auditable (every parameter traces to a documented data
source), extensible (new sources, FRPEs, or logic tree branches can be
added modularly), and applicable to any power system with operational
generation, frequency, and incident data. It establishes a bridge
between two mature probabilistic traditions, geophysical hazard
analysis and power system reliability, offering a complementary
perspective on frequency risk quantification for power systems
navigating the energy transition.

As the first application of the PSHA hazard integral to power system frequency exceedance, this work establishes a methodological foundation rather than a definitive standard. The author invites critical review, independent reproduction, and extension by the wider power systems and hazard analysis communities, with the aim of establishing best practices for probabilistic frequency hazard assessment that improve its reliability, comparability, and reproducibility across different systems and implementations. Future publications will address the application of PFHA to Future Energy Scenarios and to other synchronous areas.

%
%

\section*{Data Availability}\label{data-availability}

All datasets used in this study were obtained from publicly accessible NESO and Elexon sources. System inertia, demand, frequency response, and one-second system frequency data are available from NESO publications and the NESO Data Portal (\texttt{data.neso.energy}). Per-BMU generation data (B1610) and interconnector flow data are available through the Elexon BMRS API. GC0105 frequency event incident reports are published by NESO under Grid Code Operating Code OC6.6. No proprietary, confidential, or access-restricted data were used at any stage of the analysis.

%
%
%

\bibliographystyle{IEEEtran}
\bibliography{references}

@misc{neso_frcr_2024,
  author       = {{National Energy System Operator}},
  title        = {{Frequency Risk and Control Report 2024}},
  year         = {2025},
  month        = mar,
  institution  = {National Energy System Operator},
  address      = {Great Britain}
}

@misc{ngeso_inertia_naspi_2021,
  author       = {{National Grid ESO}},
  title        = {{System Inertia Monitoring}},
  year         = {2021},
  month        = jun,
  note         = {NASPI Webinar, 30 June 2021},
  institution  = {National Grid ESO}
}

@misc{ngeso_2019_aug9,
  author       = {{National Grid ESO}},
  title        = {{Technical Report on the Events of 9 August 2019}},
  year         = {2019},
  month        = sep,
  institution  = {National Grid ESO},
  address      = {Great Britain}
}

@misc{ofgem_frcr_2025_review,
  author       = {{Ofgem}},
  title        = {{Review of NESO's Frequency Containment Provisions and the Frequency Risk and Control Report 2025}},
  year         = {2025},
  note         = {Consultant review commissioned by Ofgem}
}

@article{cornell_1968,
  author       = {Cornell, C. A.},
  title        = {Engineering Seismic Risk Analysis},
  journal      = {Bulletin of the Seismological Society of America},
  year         = {1968},
  volume       = {58},
  number       = {5},
  pages        = {1583--1606},
  month        = oct
}

@book{baker_bradley_stafford_2021,
  author       = {Baker, Jack W. and Bradley, Brendan A. and Stafford, Peter J.},
  title        = {Seismic Hazard and Risk Analysis},
  publisher    = {Cambridge University Press},
  address      = {Cambridge, U.K.},
  year         = {2021}
}

@misc{baker_2013_psha,
  author       = {Baker, Jack W.},
  title        = {{An Introduction to Probabilistic Seismic Hazard Analysis}},
  year         = {2013},
  note         = {White paper, version 2.0.1}
}

@misc{sshac_1997,
  author       = {{Senior Seismic Hazard Analysis Committee}},
  title        = {{Recommendations for Probabilistic Seismic Hazard Analysis: Guidance on Uncertainty and Use of Experts}},
  year         = {1997},
  note         = {{NUREG/CR-6372}}
}

@article{grezio_2017_ptha,
  author       = {Grezio, A. and others},
  title        = {{Probabilistic Tsunami Hazard Analysis: Multiple Sources and Global Applications}},
  journal      = {Reviews of Geophysics},
  year         = {2017},
  volume       = {55},
  number       = {4},
  pages        = {1158--1198},
  month        = dec
}

@article{marzocchi_bebbington_2012,
  author       = {Marzocchi, W. and Bebbington, G.},
  title        = {Probabilistic Eruption Forecasting at Short and Long Time Scales},
  journal      = {Bulletin of Volcanology},
  year         = {2012},
  volume       = {74},
  number       = {8},
  pages        = {1777--1805}
}

@techreport{vlachopoulou_2016,
  author       = {Vlachopoulou, M. and others},
  title        = {{Trial Implementation of the High-Impact, Low-Frequency Power Grid Events Risk Framework}},
  institution  = {Pacific Northwest National Laboratory},
  year         = {2016},
  number       = {PNNL-25667},
  address      = {Richland, WA, USA}
}

@article{anderson_mirheydar_1990,
  author       = {Anderson, P. M. and Mirheydar, A. A.},
  title        = {A Low-Order System Frequency Response Model},
  journal      = {IEEE Transactions on Power Systems},
  year         = {1990},
  volume       = {5},
  number       = {3},
  pages        = {720--729},
  month        = aug
}

@misc{neso_eac_2024,
  author       = {{National Energy System Operator}},
  title        = {{Enduring Auction Capability (EAC) --- Market Information}},
  year         = {2024},
  institution  = {National Energy System Operator}
}

@book{kundur_1994,
  author       = {Kundur, P.},
  title        = {Power System Stability and Control},
  publisher    = {McGraw-Hill},
  address      = {New York, NY, USA},
  year         = {1994}
}

@book{billinton_allan_1996,
  author       = {Billinton, Roy and Allan, Ronald N.},
  title        = {Reliability Evaluation of Power Systems},
  edition      = {2},
  publisher    = {Plenum},
  address      = {New York, NY, USA},
  year         = {1996}
}

@book{milano_2010,
  author       = {Milano, Federico},
  title        = {Power System Modelling and Scripting},
  publisher    = {Springer},
  address      = {Berlin, Germany},
  year         = {2010}
}

@book{li_2005,
  author       = {Li, Wenyuan},
  title        = {Risk Assessment of Power Systems: Models, Methods, and Applications},
  publisher    = {IEEE Press},
  address      = {Piscataway, NJ, USA},
  year         = {2005}
}

@misc{aemo_psfrr_2022,
  author       = {{Australian Energy Market Operator}},
  title        = {{Power System Frequency Risk Review: 2022 Final Report}},
  year         = {2022},
  month        = jul,
  institution  = {AEMO}
}

@misc{aemo_gpsrr_2025,
  author       = {{Australian Energy Market Operator}},
  title        = {{General Power System Risk Review 2025}},
  year         = {2025},
  institution  = {AEMO}
}

@techreport{entsoe_inertia_2021,
  author       = {{ENTSO-E}},
  title        = {{Inertia and Rate of Change of Frequency (RoCoF)}},
  year         = {2021},
  month        = dec,
  institution  = {ENTSO-E},
  note         = {Project Inertia Phase I}
}

@techreport{entsoe_inertia_2023,
  author       = {{ENTSO-E}},
  title        = {{Project Inertia Phase II: Updated Frequency Stability Analysis in Long-Term Scenarios, Relevant Solutions and Mitigation Measures}},
  year         = {2023},
  month        = nov,
  institution  = {ENTSO-E}
}

@article{wen_bu_xin_2021,
  author       = {Wen, J. and Bu, S. and Xin, H.},
  title        = {Probabilistic Assessment on Area-Level Frequency Nadir/Vertex for Operational Planning},
  journal      = {IEEE Open Access Journal of Power and Energy},
  year         = {2021},
  volume       = {8},
  pages        = {341--351},
  doi          = {10.1109/OAJPE.2021.3108428}
}

@article{milanovic_2017,
  author       = {Milanovi\'{c}, Jovica V.},
  title        = {Probabilistic Stability Analysis: The Way Forward for Stability Analysis of Sustainable Power Systems},
  journal      = {Philosophical Transactions of the Royal Society A},
  year         = {2017},
  volume       = {375},
  number       = {2100},
  pages        = {20160296},
  doi          = {10.1098/rsta.2016.0296}
}

@article{shahzad_2022,
  author       = {Shahzad, Umair},
  title        = {A Probabilistic Framework for Power System Large-Disturbance Global Instability Risk Assessment in the Presence of Renewable Wind Generation},
  journal      = {arXiv preprint},
  year         = {2022},
  note         = {arXiv:2205.03629}
}

\end{document}